\documentclass[twocolumn]{aastex63}

\received{June 1, 2019}
\revised{January 10, 2019}
\accepted{\today}
\submitjournal{AJ}

\shorttitle{ZTF \emph{LISA} Source Survey}
\shortauthors{Burdge et al.}
\graphicspath{{./}{figures/}}

\begin{document}

\title{A systematic search of Zwicky Transient Facility data for ultracompact binary \emph{LISA}-detectable gravitational-wave sources}

\correspondingauthor{Kevin B. Burdge}
\email{kburdge@caltech.edu}

\author[0000-0002-7226-836X]{Kevin B. Burdge}
\affiliation{Division of Physics, Mathematics and Astronomy, California Institute of Technology, Pasadena, CA 91125, USA}
\author[0000-0002-8850-3627]{Thomas A. Prince}
\affiliation{Division of Physics, Mathematics and Astronomy, California Institute of Technology, Pasadena, CA 91125, USA}
\author[0000-0002-4544-0750]{Jim Fuller}
\affiliation{Division of Physics, Mathematics and Astronomy, California Institute of Technology, Pasadena, CA 91125, USA}
\author[0000-0001-6295-2881]{David L.\ Kaplan}
\affiliation{Department of Physics, University of Wisconsin-Milwaukee, Milwaukee, WI, USA}
\author[0000-0002-2498-7589]{Thomas R. Marsh}
\affiliation{Department of Physics, University of Warwick, Coventry, CV4 7AL, UK}
\author[0000-0001-9873-0121]{Pier-Emmanuel Tremblay}
\affiliation{Department of Physics, University of Warwick, Coventry, CV4 7AL, UK}
\author[0000-0002-1945-2299]{Zhuyun Zhuang}
\affiliation{Division of Physics, Mathematics and Astronomy, California Institute of Technology, Pasadena, CA 91125, USA}

\author[0000-0001-8018-5348]{Eric C. Bellm}
\affiliation{DIRAC Institute, Department of Astronomy, University of Washington, 3910 15th Avenue NE, Seattle, WA 98195, USA}
\author[0000-0002-4770-5388]{Ilaria Caiazzo}
\affiliation{Division of Physics, Mathematics and Astronomy, California Institute of Technology, Pasadena, CA 91125, USA}
\author[0000-0002-8262-2924]{Michael W. Coughlin}
\affiliation{School of Physics and Astronomy, University of Minnesota, Minneapolis, Minnesota 55455, USA}
\author[0000-0003-4236-9642]{Vik S. Dhillon}
\affiliation{Department of Physics \& Astronomy, University of Sheffield, Sheffield S3 7RH, UK}
\affiliation{Instituto de Astrof\'\i sica de Canarias, V\'\i a L\'actea s/n, La Laguna, E-38205 Tenerife, Spain}
\author[0000-0002-2761-3005]{Boris Gaensicke}
\affiliation{Department of Physics, University of Warwick, Coventry, CV4 7AL, UK}
\author[0000-0002-4717-5102]{Pablo Rodr\'\i guez-Gil}
\affiliation{Instituto de Astrof\'\i sica de Canarias, V\'\i a L\'actea s/n, La Laguna, E-38205 Tenerife, Spain}
\affiliation{Departamento de Astrof\'\i sica, Universidad de La Laguna, E-38206 La Laguna, Tenerife, Spain}
\author[0000-0002-3168-0139]{Matthew J. Graham}
\affiliation{Division of Physics, Mathematics and Astronomy, California Institute of Technology, Pasadena, CA 91125, USA}
\author[0000-0001-5941-2286]{JJ Hermes}
\affiliation{Department of Astronomy, Boston University, 725 Commonwealth Ave., Boston, MA 02215, USA}
\author[0000-0002-6540-1484]{Thomas Kupfer}
\affiliation{Kavli Institute for Theoretical Physics, University of California, Santa Barbara, CA 93106, USA}
\author[0000-0001-7221-855X]{S. P. Littlefair}
\affiliation{Department of Physics \& Astronomy, University of Sheffield, Sheffield S3 7RH, UK}
\author[0000-0001-7016-1692]{Przemek Mr{\'o}z}
\affiliation{Division of Physics, Mathematics and Astronomy, California Institute of Technology, Pasadena, CA 91125, USA}
\author[0000-0002-9656-4032]{E.~S. Phinney}
\affiliation{Theoretical Astrophysics, 350-17 California Institute of Technology, Pasadena CA 91125, USA}
\author[0000-0002-2626-2872]{Jan van~Roestel}
\affiliation{Division of Physics, Mathematics and Astronomy, California Institute of Technology, Pasadena, CA 91125, USA}
\author[0000-0001-6747-8509]{Yuhan Yao}
\affiliation{Division of Physics, Mathematics and Astronomy, California Institute of Technology, Pasadena, CA 91125, USA}

\author[0000-0002-5884-7867]{Richard G. Dekany}
\affiliation{Caltech Optical Observatories, California Institute of Technology, Pasadena, CA, USA}
\author{Andrew J. Drake}
\affiliation{Division of Physics, Mathematics and Astronomy, California Institute of Technology, Pasadena, CA 91125, USA}
\author[0000-0001-5060-8733]{Dmitry A. Duev}
\affiliation{Division of Physics, Mathematics and Astronomy, California Institute of Technology, Pasadena, CA 91125, USA}
\author{David Hale}
\affiliation{Caltech Optical Observatories, California Institute of Technology, Pasadena, CA, USA}
\author{Michael Feeney}
\affiliation{Caltech Optical Observatories, California Institute of Technology, Pasadena, CA, USA}
\author[0000-0003-3367-3415]{George Helou}
\affiliation{IPAC, California Institute of Technology, 1200 E. California Blvd, Pasadena, CA 91125, USA}
\author{Stephen Kaye}
\affiliation{Caltech Optical Observatories, California Institute of Technology, Pasadena, CA, USA}
\author[0000-0003-2242-0244]{Ashish.~A.~Mahabal}
\affiliation{Division of Physics, Mathematics and Astronomy, California Institute of Technology, Pasadena, CA 91125, USA}
\affiliation{Center for Data Driven Discovery, California Institute of Technology, Pasadena, CA 91125, USA}
\author[0000-0002-8532-9395]{Frank J. Masci}
\affiliation{IPAC, California Institute of Technology, 1200 E. California Blvd, Pasadena, CA 91125, USA}
\author[0000-0002-0387-370X]{Reed Riddle}
\affiliation{Caltech Optical Observatories, California Institute of Technology, Pasadena, CA, USA}
\author[0000-0001-7062-9726]{Roger Smith}
\affiliation{Caltech Optical Observatories, California Institute of Technology, Pasadena, CA, USA}
\author[0000-0001-6753-1488]{Maayane T. Soumagnac}
\affiliation{Lawrence Berkeley National Laboratory, 1 Cyclotron Road, Berkeley, CA 94720, USA}
\affiliation{Department of Particle Physics and Astrophysics, Weizmann Institute of Science, Rehovot 76100, Israel}
\author[0000-0001-5390-8563]{S. R. Kulkarni}
\affiliation{Division of Physics, Mathematics and Astronomy, California Institute of Technology, Pasadena, CA 91125, USA}



\begin{abstract}

Using photometry collected with the Zwicky Transient Facility (ZTF), we are conducting an ongoing survey for binary systems with short orbital periods ($P_{\rm b}<1\rm \,hr)$ with the goal of identifying new gravitational-wave sources detectable by the upcoming \emph{Laser Interferometer Space Antenna} (\emph{LISA}). Here, we present a sample of fifteen binary systems discovered thus far, with orbital periods ranging from $6.91\rm\,min$ to $56.35\rm\,min$. Of the fifteen systems, seven are eclipsing systems which do not show signs of significant mass transfer. Additionally, we have discovered two AM Canum Venaticorum (AM CVn) systems and six systems exhibiting primarily ellipsoidal variations in their light curves. We present follow-up spectroscopy and high-speed photometry confirming the nature of these systems, estimates of their \emph{LISA} signal-to-noise ratios (SNR), and a discussion of their physical characteristics.

\end{abstract}



\section{Introduction} \label{sec:intro}

The upcoming \emph{Laser Interferometer Space Antenna} (\emph{LISA}; \citet{Amaro-Seoane2017}) will be a space-based millihertz-frequency gravitational-wave detector. Astrophysical sources of gravitational radiation in this frequency band include merging supermassive black hole binaries, extreme mass ratio inspirals, and Galactic binaries with short orbital periods. With an anticipated number of more than ten thousand detectable sources in the Milky Way \citep{Nissanke2012}, close double-degenerate binaries are by far the largest population of detectable gravitational-wave sources, but only a few have been discovered to date \citep{Kupfer2018}. This population of sources presents a unique opportunity to use the synergy of information carried by electromagnetic and gravitational radiation to understand processes such as binary evolution, the population of Type Ia supernova and R Coronae Borealis (R CrB) progenitors \citep{Webbink1984}, tidal physics in degenerate objects \citep{Fuller2012}, and Galactic structure \citep{Korol2019}; however, because these sources are so numerous, they will also present a technical challenge by acting as a formidable background for \emph{LISA} \citep{Nelemans2001}.

Since the beginning of science operations in March 2018, Zwicky Transient Facility (ZTF) \citep{Bellm2019,Dekany2020,Graham2019,Masci2019} has accumulated hundreds of epochs across the northern sky, with more than a thousand in some regions. Here, we present results from a new survey using ZTF data with the goal of discovering \emph{LISA} sources in the optical time domain. The survey identifies objects which undergo periodic flux variations on short timescales. We conduct additional spectroscopic and photometric follow-up of objects which exhibit a strong periodic signal with a period shorter than $30 \rm \, min$. By targeting objects in this manner in an all-sky survey, we can search millions of candidates, in contrast to the narrower selection criteria used in surveys such as the highly successful Extremely Low Mass (ELM) survey. The ELM survey spectroscopically followed up all candidates in a narrow parameter space, resulting in the discovery of 98 detached double-white dwarfs (DWDs), over half of the known double-degenerate population \citep{ELMI,ELMII,ELMIII,ELMIV,ELMV,ELMVI,ELMVII,ELMVIII}, including several systems which are strong candidate \emph{LISA}-detectable gravitational-wave sources \citep{Brown2011,Kilic2014,Brown2020b}. Until recently, there were only two known detached eclipsing binaries under an hour orbital period, both discoveries made by the ELM survey \citep{Brown2011,Brown2017}. 

 In June 2019, using ZTF's first internal data release, we discovered the shortest orbital period eclipsing binary system known, ZTF J1539+5027, with an orbital period of just $6.91\rm\,min$ \citep{Burdge2019a}. We originally tested the viability of a photometric selection strategy using archival Palomar Transient Factory (PTF) data \citep{Law2009}, and discovered PTF J0533+0209, a $\approx 20.6$ minute orbital period detached DBA WD binary \citep{Burdge2019b}. Since these two discoveries, we have confirmed 13 additional short period binary systems either spectroscopically via radial velocity shifts, or photometrically if they exhibit eclipses (an unambiguous indicator of binarity). These systems are diverse, and include detached DWDs, accreting AM Canum Venaticorum (AM CVn) objects \citep{Ramsay2018}, and accreting systems involving helium stars/hot sub-luminous subdwarfs (sdBs/sdOs) \citep{Heber2016}.  Even among just the detached DWD population, there exists a rich phase space of possible combinations of core compositions involving helium-core white dwarfs (He WD) and carbon-oxygen core WDs (CO WD), each of which originate from distinct evolutionary scenarios. Thus, in this work, we devote some discussion to describing the unique characteristics of each system. As a point of reference to guide the reader, we have included Table \ref{tab:objects} below to highlight essential characteristics of the fifteen systems discussed in this work.

\begin{deluxetable*}{cccccc}
\tablenum{1}
\tablecaption{ZTF/PTF Short period binaries \label{tab:objects}}
\tablewidth{0pt}
\tablehead{
\colhead{Name} & \colhead{Right Ascension}  & \colhead{Declination} &
\colhead{Orbital period} & \colhead{Nature of photometric variability} & \colhead{Spectroscopic characteristics} \\
\colhead{} & \colhead{(h:m:s)} & \colhead{(d:m:s)}  & \colhead{(min)} &
\colhead{} & \colhead{} 
}
\startdata
ZTF J1539+5027$^1$ & 15:39:32.16   & +50:27:38.72 & \phn6.91 & Eclipsing+irradiation & DA, double-lined \\
ZTF J0538+1953 & 05:38:02.73   & +19:53:02.89 & 14.44 & Eclipsing+irradiation &  DA, double-lined \\
ZTF J1905+3134 &  19:05:11.34  & +31:34:32.37 & 17.20 & Eclipsing High State AM CVn & Double-peaked \ion{He}{2} emission \\
PTF J0533+0209$^2$ &  05:33:32.06  & +02:09:11.51 & 20.57 & Ellipsoidal & DBA single-lined \\
ZTF J2029+1534 &  20:29:22.31  & +15:34:30.97 & 20.87 & Eclipsing & DA, double-lined \\
ZTF J0722$-$1839 &  07:22:21.49  & -18:39:30.57 & 23.70 & Eclipsing & DA, double-lined \\
ZTF J1749+0924 &  17:49:55.30  & +09:24:32.40 & 26.43 & Eclipsing & DA, double-lined \\
ZTF J2228+4949 &  22:28:27.07  & +49:49:16.44 & 28.56 & High State AM CVn & Double-peaked \ion{He}{2} emission \\
ZTF J1946+3203 &  19:46:03.89  & +32:03:13.13 & 33.56 & Eclipsing+Ellipsoidal & DAB/sdB, single-lined \\
ZTF J0643+0318 &  06:43:36.77  & +03:18:27.45 & 36.91 & Accreting He star & \ion{He}{1} absorption/\ion{He}{2} emission \\
ZTF J0640+1738 &  06:40:18.69  & +17:38:45.01 & 37.27 & Ellipsoidal & sdB, single-lined \\
ZTF J2130+4420$^3$ &  21:30:56.71  & +44:20:46.42 & 39.34 & Ellipsoidal & sdB, single-lined \\
ZTF J1901+5309$^4$ &  19:01:25.42  & +53:09:29.27 & 40.60 & Eclipsing & DA, double-lined \\
ZTF J2320+3750 &  23:20:20.43  & +37:50:30.84 & 55.25 & Ellipsoidal & DA, single-lined \\
ZTF J2055+4651$^5$ &  20:55:15.98  & +46:51:06.45 & 56.35 & Eclipsing+Ellipsoidal & sdB, single-lined \\
\enddata
\tablecomments{The coordinates and basic photometric and spectroscopic characteristics of the fifteen short period binaries discovered so far using PTF/ZTF data. Coordinates are taken from \emph{Gaia} and are in J2000.0. For apparent magnitudes, see Table \ref{tab:phot}. More precise orbital periods and uncertainties are reported in Table \ref{tab:measured_quantities}. \tablerefs{\citet{Burdge2019a}$^1$, \citet{Burdge2019b}$^2$, \citet{Kupfer2020a}$^3$, \citet{Coughlin2019}$^4$, \citet{Kupfer2020b}$^5$}}
\end{deluxetable*}

\section{Methods} \label{sec:methods}

\subsection{Sample Selection}

We used the Pan-STARRS1 (PS1) source catalog \citep{Chambers2016} to select our sample of objects to period search. Focusing on ``blue'' objects, we imposed a photometric color selection of $(g-r)<0.2$ and additionally $(r-i)<0.2$ (see Table \ref{tab:phot}). We selected blue objects, because we expected that close binary WDs, especially those at orbital periods sufficiently short to be strong \emph{LISA} sources, should have elevated temperatures due to tidal heating \citep{Fuller2013}.

\begin{deluxetable*}{ccccccc}[htbp]
\tablenum{2}
\tablecaption{Optical and ultraviolet apparent magnitudes of the sample \label{tab:phot}}
\tablewidth{0pt}
\tablehead{\colhead{Survey:} & \colhead{GALEX}  & \colhead{GALEX} & \colhead{Pan-STARRS1} & \colhead{Pan-STARRS1} & \colhead{Pan-STARRS1} & \colhead{\emph{Gaia}}  \\
\colhead{Filter:} & \colhead{FUV}  & \colhead{NUV} & \colhead{$g$} & \colhead{$r$} & \colhead{$i$} & \colhead{$G$}  \\
 & \colhead{($\rm m_{AB}$)} & \colhead{($\rm m_{AB}$)}  & \colhead{($\rm m_{AB}$)}  & \colhead{($\rm m_{AB}$)} & \colhead{($\rm m_{AB}$)} & \colhead{($\rm m_{V}$)} 
}

\startdata
ZTF J1539+5027 & $18.754\pm0.084$  & $19.334\pm0.073$ & $20.134\pm0.041$ & $20.521\pm0.069$	 & $20.805\pm0.049$ & $20.491\pm0.017$ \\
ZTF J0538+1953 &   & $19.59\pm0.16$ & $18.777\pm0.015$ & $18.881\pm0.011$	 & $19.1049\pm0.0060$ & $18.8346\pm0.0061$ \\
ZTF J1905+3134 &   &  & $20.778\pm0.046$ & $21.15\pm0.28$	 & $21.19\pm0.13$ & $20.900\pm0.030$ \\
PTF J0533+0209 &   & $20.38\pm0.28$ & $18.995\pm0.015$ & $19.150\pm0.013$	 & $19.405\pm0.014$ & $19.1030\pm0.0075$\\
ZTF J2029+1534 &  $20.12\pm0.23$ & $20.49\pm0.20$ & $20.380\pm0.024$ & $20.662\pm0.047$	 & $20.962\pm0.093$ & $20.5745\pm0.0118$\\
ZTF J0722$-$1839 &  &  & $18.976\pm0.012$ & $19.194\pm0.016$ & $19.4888\pm0.0095$ & $19.1040\pm0.0061$\\
ZTF J1749+0924 & $20.19\pm0.26$ &  & $20.357\pm0.022$ & $20.474\pm0.014$ & $20.666\pm0.029$ & $20.538\pm0.013$\\
ZTF J2228+4949 &   & $19.43\pm0.12$ & $19.317\pm0.028$ & $19.390\pm0.034$ & $19.571\pm0.022$ & $19.2758\pm0.0060$\\
ZTF J1946+3203 &   & & $19.400\pm0.035$ & $19.266\pm0.020$ & $19.240\pm0.025$ & $19.1905\pm0.0059$\\
ZTF J0643+0318 &   & $21.68\pm0.48$ & $18.384\pm0.012$ & $18.269\pm0.058$ & $18.306\pm0.037$ & $18.271\pm0.019$ \\
ZTF J0640+1738 &   &  & $19.222\pm0.017$ & $19.409\pm0.024$ & $19.486\pm0.016$ & $19.2521\pm0.0062$ \\
ZTF J2130+4420 &  & $15.727\pm0.012$ & $15.3269\pm0.0049$ & $15.6412\pm0.0078$ & $15.806\pm0.028$ & $15.4611\pm0.0061$  \\
ZTF J1901+5309 &   & $17.331\pm0.041$ & $17.8909\pm0.0055$ & $18.2512\pm0.0031$ & $18.5764\pm0.0055$ & $18.0686\pm0.0023$ \\
ZTF J2320+3750  &   & $21.39\pm0.28$ & $19.411\pm0.024$ & $19.394\pm0.025$ & $19.536\pm0.028$ & $19.3997\pm0.0070$ \\
ZTF J2055+4651 &  &  & $17.720\pm0.042$ & $17.652\pm0.023$ & $17.647\pm0.015$ & $17.6539\pm0.0093$  \\
\enddata

\end{deluxetable*}

\subsection{ZTF Photometry}

We used ZTF photometry to identify all of our sources (with the exception of PTF J0533+0209). ZTF is a northern sky synoptic survey based on the 48-inch Samuel Oschin Schmidt telescope at Palomar Observatory, surveying the sky in ZTF $g$-, $r$-, and $i$-bands down to a declination of $-28^{\circ}$ with $30\rm\,s$ exposures. ZTF has a 47 $\rm deg^2$ field of view and a median $5 \sigma$ limiting magnitude of 20.8 in $g$-band, 20.6 in $r$-band, and 20.2 in $i$-band. The instrument's large field of view allows it to rapidly accumulate a large number of epochs across the sky, and the resulting dense sampling serves as a crucial element in identifying sources.

\subsection{Period Finding}

Because the fidelity of period finding depends strongly on the number of samples in a lightcurve (especially at short periods where there are large numbers of trial frequencies), rather than impose a cutoff on apparent magnitude, we instead required a minimum of 50 total photometric $5 \sigma$ detections in archival ZTF data. This selection produced approximately $10$ million candidates to search. We combined data from multiple filters by computing the median magnitude in each filter, and shifting $g$- and $i$-band so that their median magnitude matched the $r$-band data. We do this in order to maximize the number of epochs available in each lightcurve.
We used a graphics processing unit (GPU) implementation of the conditional entropy period finding algorithm \citep{Graham2013} and later the Box Least Squares (BLS) algorithm \citep{Kovacs2002}. Although BLS is optimized for identifying eclipsing systems, conditional entropy is computationally less costly, and was responsible for the discovery of 12 of the 15 systems in the sample. The only system discovered in a later BLS search was ZTF J1749+0924. We originally discovered ZTF J2130+4420, which is the brightest object in the sample, using the analysis of variance algorithm \citep{Schwarzenberg1996} in a search of bright sources between the WD track and the main sequence. It was later recovered by both the conditional entropy algorithm, and as part of a search of the hot subdwarf catalog \citep{Geier2017} as discussed in \citet{Kupfer2020a} (PTF J0533+0209 was also discovered with the AOV algorithm--see \citet{Burdge2019b}). The BLS algorithm is able to recover all systems as well, but with greater than an order of magnitude longer run-time relative to conditional entropy. The conditional entropy algorithm phase-folds lightcurves on a grid of trial frequencies, partitions the lightcurve into magnitude and phase bins, and computes the conditional entropy of the partitioned phase-folded lightcurve corresponding to each frequency. For further details on the sensitivity of this algorithm, please see the Appendix.

We determine the best period of the object by selecting the trial frequency which minimizes the conditional entropy. We visually inspect the most significant phase-folded lightcurves (all with significance greater than 8), where significance is simply defined as the minimum entropy value divided by the standard deviation of entropy values across the full periodogram, and from this we select the most promising candidates for follow-up. We performed this visual inspection on 24635 lightcurves with a minimum entropy corresponding to a period below 30 minutes, and determined that 337 of these were likely real periodic signals based on the appearance of their lightcurves. We then began to systematically follow-up systems which exhibited photometric behavior suggestive of binarity (eclipses, or asymmetric minima in ellipsoidal variables). Follow-up of objects in this sample is ongoing.

One challenge of searching for periodic variables in a sample of millions of objects over a large frequency grid (for further details see the appendix) is the increase in opportunity for objects to exhibit significant random fluctuations in their power spectra. This means that many of the lower amplitude periodic variables recovered actually show a smaller ``significance'' than some non-periodic objects, hence the value in visually inspecting a moderate number of objects at low significance, rather than only the most significant sources. Note that targeting objects with ``best'' periods under 30 mins yielded many discoveries with orbital periods between 30 and 60 minutes, as systems which exhibit ellipsoidal variations and eclipsing systems in which there is a prominent secondary eclipse exhibit significant power at twice their orbital frequency.

\begin{figure*}[ht]
\begin{center}
\includegraphics[width=1.0\textwidth]{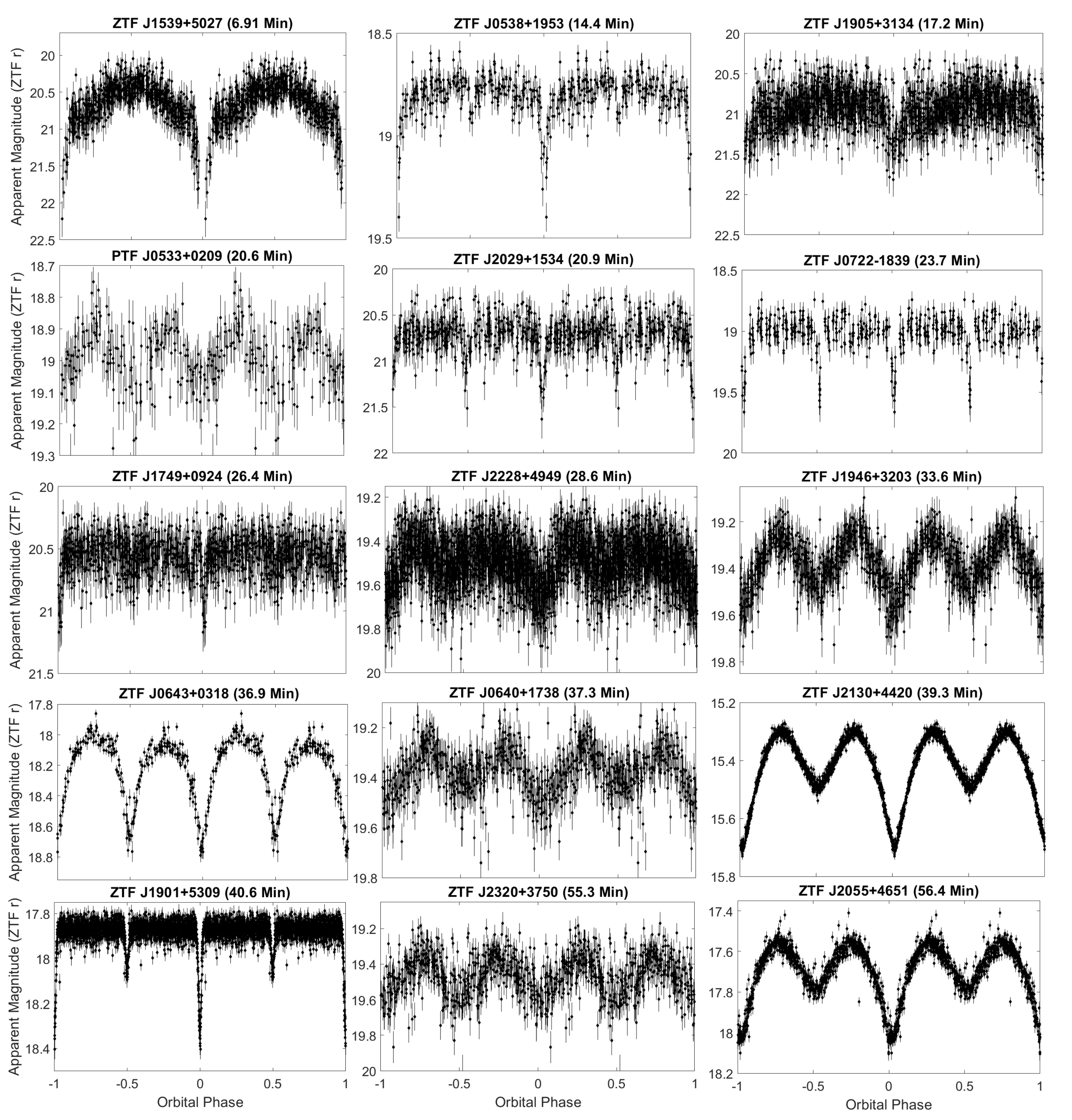}
\caption{The phase-folded ZTF lightcurves of the sample (all filters combined), at periods as recovered by the algorithm described in Section 2.3.}
\label{fig:ZTF}
\end{center}
\end{figure*}

The ZTF phase-folded lightcurves of all short period binary systems discovered from this sample are illustrated in Figure \ref{fig:ZTF}.

In order to determine the uncertainties in the orbital period reported in Table \ref{tab:measured_quantities}, we used a bootstrapping technique in which we sub-sampled each ZTF lightcurve 1000 times, using 90 percent of the points in each sub-sample, with replacement, and computed the standard deviation of the best periods of the 1000 trials. For ellipsoidal systems, we used Lomb-Scargle \citep{Lomb1976,Scargle1982} for this bootstrapping exercise, and for eclipsing systems, we used BLS. For ZTF J1539+5027 and PTF J0533+0209, which have well-measured orbital evolution, we instead used the linear component of the quadratic fit to the timing residuals to determine the uncertainty in the period (for further details, see \citet{Burdge2019a}).

\begin{figure*}[htpb]
\begin{center}
\includegraphics[width=1.0\textwidth]{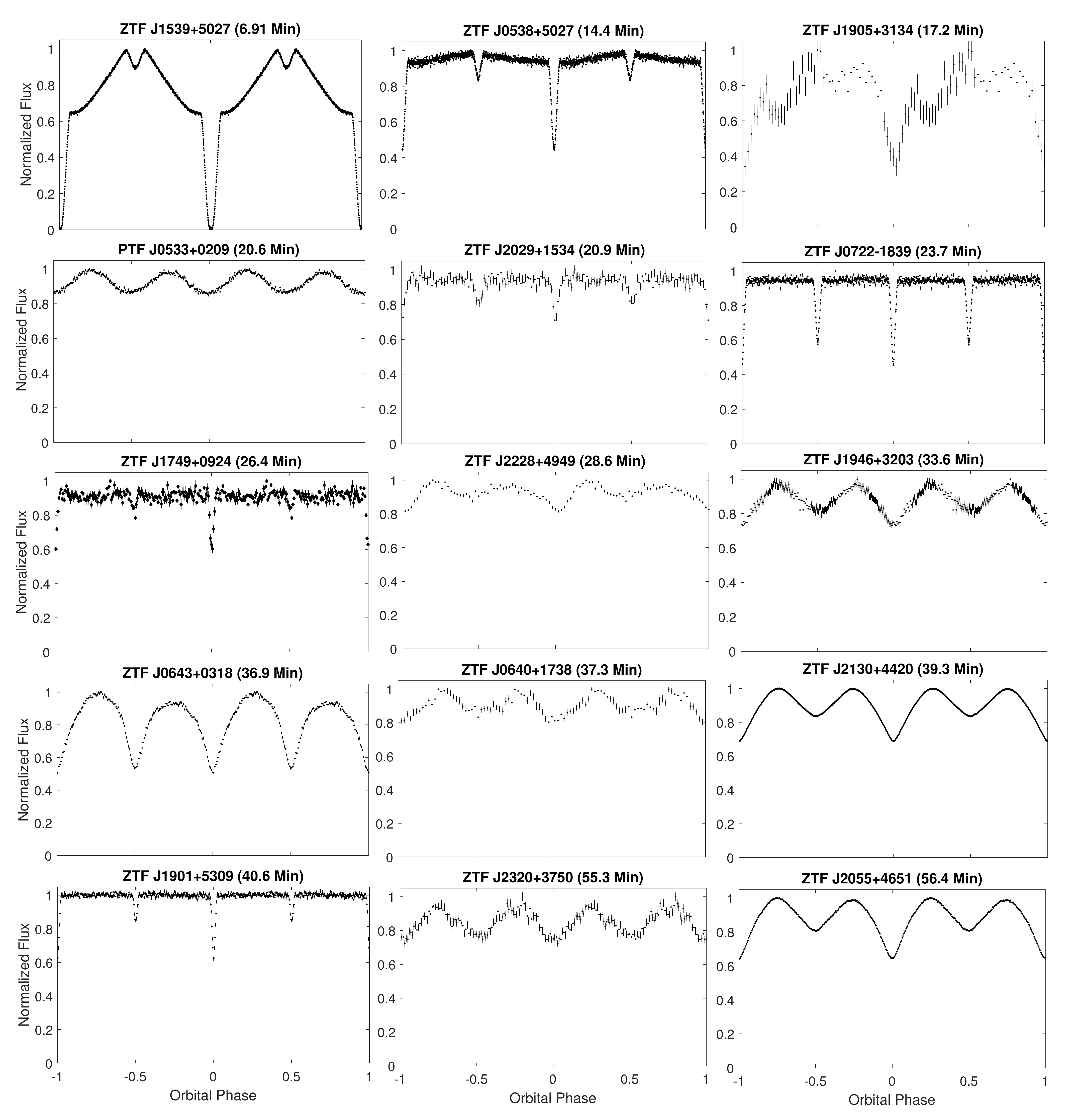}
\caption{The best available lightcurves for our sample, phase-binned at the orbital periods. See Table \ref{tab:Observations} for further details on which instruments were used to obtain each lightcurve. J2228+4949 and J0640+1738 have only ZTF lightcurves, and thus we plot the binned versions of those here. All of these systems were confirmed as binaries via spectroscopic follow-up.}
\label{fig:FollowUpPhot}
\end{center}
\end{figure*}  

\begin{figure*}[htp]
\begin{center}
\includegraphics[width=1.0\textwidth]{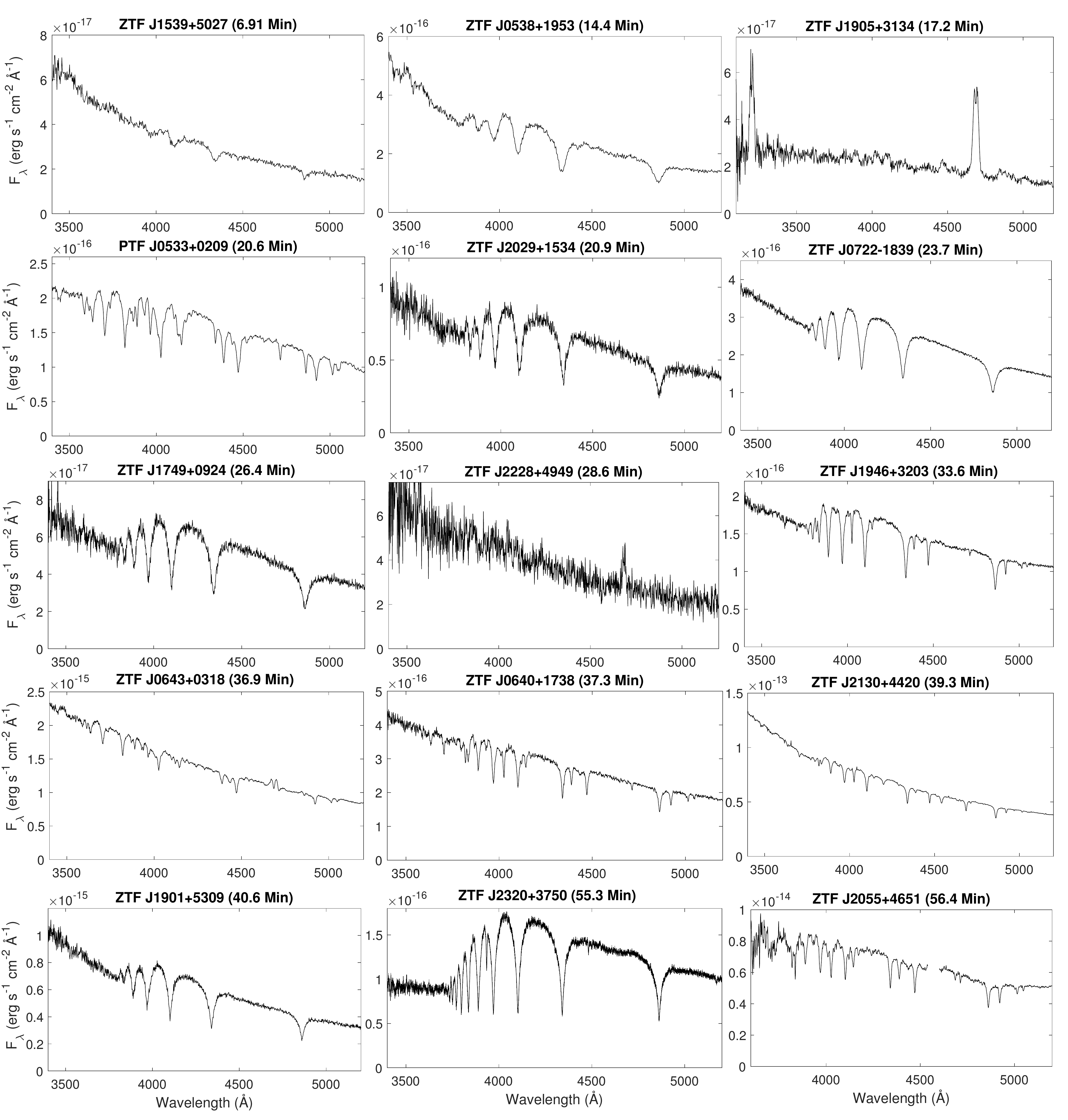}
\caption{The best available spectra for our sample, coadded in the rest frame of absorption lines. By fitting these spectra, we were able to obtain radial velocity semi-amplitudes for seven systems (see Figure \ref{fig:RVs}), and fit objects for an effective temperature and surface gravity using spectroscopic WD models. We used these spectroscopically derived parameters in combination with lightcurve modelling to derive the physical parameters reported in Table \ref{tab:derived_quantities}. See the Appendix for further details on which instruments were used to obtain each spectrum. ZTF J1539+5027, J0538+1953, J2029+1534, J0722$-$1839, J1749+0924, J1901+5309, and J2320+3750 show hydrogen-rich atmospheres, while J0533+0209, J1946+3203, J0640+1738, J2130+4420, and J2055+4651 exhibit mixed H/He atmospheres, and the three mass transferring systems, J1905+3134, J2228+4949, and J0643+0318 have only helium lines in their spectra.}
\label{fig:FollowUpSpec}
\end{center}
\end{figure*}  

Of the short period candidates recovered by conditional entropy, we immediately identified two as ellipsoidal variables with periods between 30 and 60 minutes (ZTF J2130+4420 and ZTF J2055+4651) because they clearly exhibited significant power at twice their orbital frequency, distinguishing them from the large number of $\delta$ Scuti stars in this period range \citep{Breger2000}. Spectroscopic characterization later revealed additional ellipsoidal variables with orbital periods between 30 and 60 minutes. Some candidates exhibiting a dominant period below 30 min failed to show indications of binarity in follow-up characterization. Some of these were identified as likely belonging to other classes of short period variables including pulsating WDs such as GW Vir, V777 Her, and ZZ Ceti stars \citep{Fontaine2008}, hot subdwarf p-mode pulsators \citep{Charpinet1996}, blue large amplitude pulsators \citep{Piertrukowicz2017}, He WD radial mode pulsators \citep{Kupfer2019}, rapidly rotating magnetic WDs \citep{Brinkworth2013,Reding2020}. One significant source of interlopers in the sample originates from intermediate polars \citep{Patterson1994}, which contain magnetic WDs which have been spun up to a short spin period, and exhibit periodicity on the timescale of minutes. These objects often have blue color, emit X-rays, and can show strong photometric periodicity on minute timescales due to flux variations on the WD spin period \citep{Ramsay2008}. The crucial diagnostic we use to rule out a minutes-timescale orbital period in these systems is phase-resolved spectroscopy, as they do not exhibit radial velocity shifts on the WD spin period.

\subsection{High Speed Photometry}

  High-speed photometric follow-up proved a crucial element in characterizing candidates from our period searches, as such observations can confirm the presence of eclipses and more subtle photometric effects such as Doppler beaming and gravity darkening. In this work, we present data collected from four different high speed photometers: the Caltech HIgh-speed Multi-color camERA (CHIMERA) on the 5.1 meter Hale telescope at Palomar observatory \citep{Harding2016}, the Kitt Peak Electron Multiplying CCD Demonstrator (KPED) on the 2.1 meter telescope at Kitt Peak National observatory \citep{Coughlin2019}, HiPERCAM on the 10.4 meter Gran Telescopio Canarias (\citealt{Dhillon2018}, Dhillon et al. in prep), and ULTRACAM on the 3.5 meter New Technology Telescope at La Silla observatory \citep{Dhillon2007}. In all cases, the instruments were run in frame transfer mode, effectively eliminating readout time overheads. We reduced the CHIMERA, KPED, HiPERCAM and ULTRACAM data with publically available pipelines \footnote{\url{https://github.com/mcoughlin/kp84}}$^{,}$\footnote{\url{https://github.com/HiPERCAM}}$^{,}$\footnote{\url{https://github.com/trmrsh/cpp-ultracam}}. Lightcurves obtained with these instruments are illustrated in Figure \ref{fig:FollowUpPhot}. For further details on these observations, please see the Appendix.

\subsection{Spectroscopic follow-up}

We conducted most of our spectroscopy using the Low Resolution Imaging Spectrometer (LRIS) \citep{Oke1995} on the 10-m W.\ M.\ Keck I Telescope on Mauna Kea. Exposure times varied from object to object, but were restricted to be no more than an eighth of the orbital period in order to minimize Doppler smearing. We reduced all data with the lpipe pipeline \citep{Perley2019}. For all sequences of exposures, we took a HeNeArCdZn arc at the telescope position of the object at the completion of the sequence in order to ensure stable wavelength calibration (which can depend strongly on instrument flexure). For any sequences of observations exceeding an hour, we obtained an arc at the start and finish of observations, or if longer than two hours, an arc once every two hours. For further details on instrument configuration, and other spectrographs used, please see the Appendix. The spectra we obtained are presented in Figure \ref{fig:FollowUpSpec}.

\section{Analysis and Results} \label{sec:analaysisresults}

\subsection{Lightcurve Modelling}

We used the LCURVE code \citep{Copperwheat2010} to model and infer physical parameters from the lightcurves we obtained with high speed photometers. For lightcurves with multiple bands, we simultaneously modelled all bands (measuring the relative surface brightnesses of objects at multiple wavelengths aids in determining the relative temperatures of the objects in the system). For all objects, we used the following set of free parameters: the mass ratio $q=\frac{M_{B}}{M_{A}}$, the inclination $i$, the radius of the primary component, $R_A$, the radius of the secondary component, $R_B$, the temperature of the secondary component, $T_B$, the mid-eclipse time, $t_0$, and an absorption coefficient $\alpha$ to account for effects in which the cooler component in the binary is irradiated and reprocesses flux originating from the hotter component. We fit for a different $\alpha$ for each band, as radiation reprocessing is wavelength dependent. 
Measuring the mass ratio, $q$, from lightcurve modelling arises primarily from fitting ellipsoidal modulation due to the tidal deformation of one or both components in the system. The fractional amplitude of ellipsoidal modulation, $\frac{\Delta F_{\rm ellipsoidal}}{F}$, is given by: 
\begin{equation} \label{eq:Ellipsoidal}
\frac{\Delta F_{\rm ellipsoidal}}{F}=0.15\frac{(15+u)(1+\tau)}{3-u}\left(\frac{R}{a}\right)^3q\sin^2(i),
\end{equation}
\citep{Morris1985},
where $u$ is the passband dependent linear limb darkening coefficient, and $\tau$ is the passband dependent gravity darkening coefficient. We obtain all gravity and limb darkening coefficients from \citet{Claret2020}, and Doppler beaming coefficients from \citet{Claret2020b}. For our full modelling, we use the 4 parameter Claret limb-darkening law described in \citet{Claret2000}. Previous work such as that presented in \citet{Bloemen2012} found tension between the mass ratio inferred from purely ellipsoidal modulation, and that determined by other less model-dependent measurements such as radial velocity semi-amplitudes, which they attributed to the possibility that some close binaries may not be entirely synchronized. At these short orbital periods, we expect some of our systems to be synchronized, but those at longer periods may not be \citep{Fuller2012}. In any case, we emphasize that mass ratios derived from ellipsoidal modulation come from model-dependent expressions, and thus there is value in obtaining more model-independent measurements in the form of a radial velocity semi-amplitudes as well as orbital decay due to general relativity \citep{Burdge2019a}. 

We estimated masses for five of the eclipsing systems (ZTF J0538+1953, ZTF J2029+1534, ZTF J0722$-$1839, ZTF J1749+0924, and ZTF J1901+5309) using temperatures derived from spectroscopy, and radii derived from lightcurve modelling. As discussed above, lightcurve modelling allows us to infer the radii of the two component objects with respect to the semi-major axis in these systems. We combined these measurements (as well as our spectroscopic temperature constraints) with the semi-empirical mass-radius relations described in \citet{Soares2017} in order to estimate the masses of both components. We adopted a 10 percent model error in estimating our uncertainties based on the scatter of measured radii vs estimated radii reported in \citet{Soares2017}, because mass-radius relations (especially for low mass He WDs) are sensitive to additional degrees of freedom such as the mass of the hydrogen envelope on the surface of the WD and the evolutionary history of the object (in particular, the occurrence of hydrogen shell flashes; \citet{Istrate2016}). In the future, as we measure orbital decay rates for these systems, we will be able to constrain masses in a more model-independent manner. 

\subsection{Orbital Dynamics}

\begin{figure*}[ht]
\begin{center}
\includegraphics[width=1.0\textwidth]{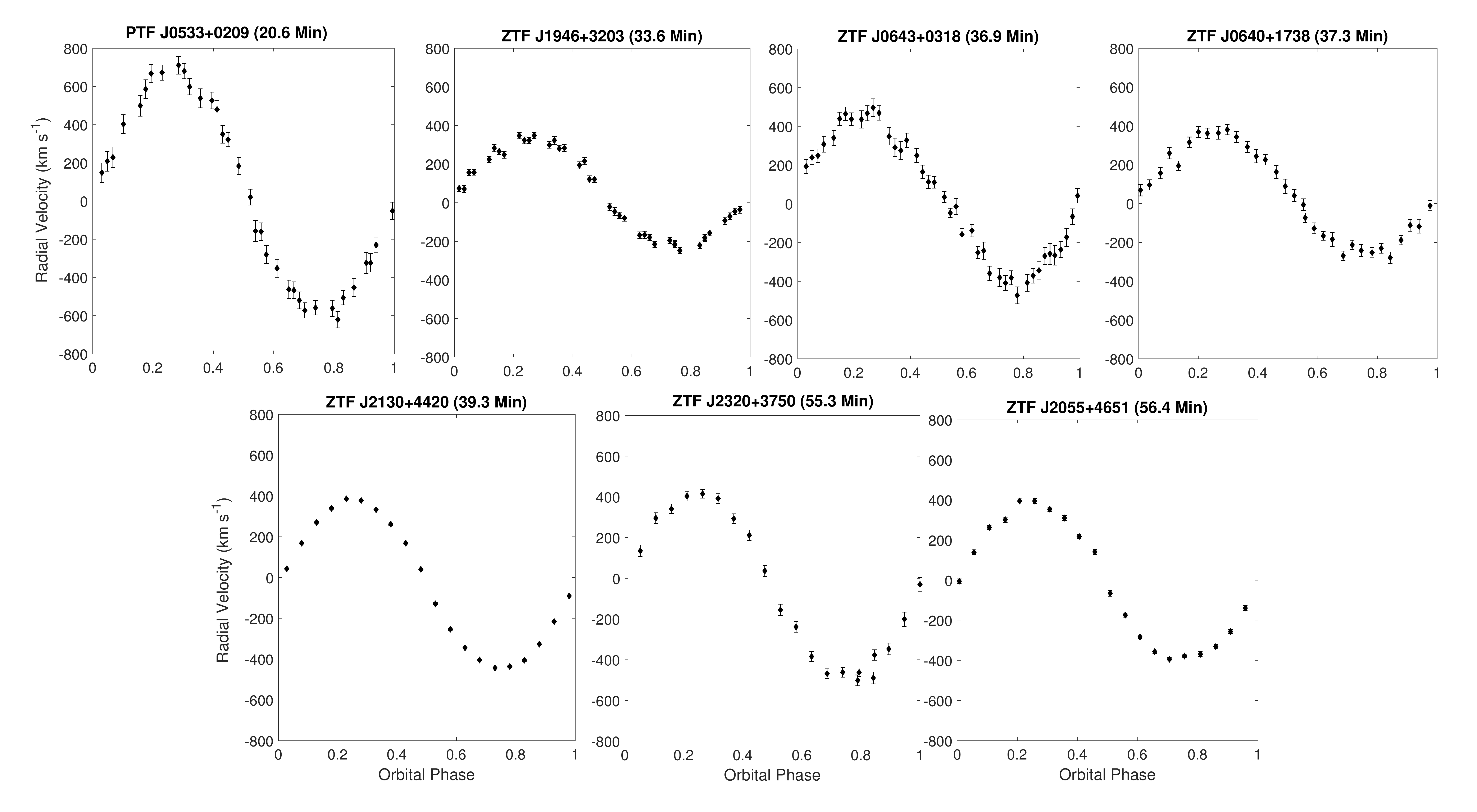}
\caption{The phase-folded radial velocity measurements for a subset of the sample which is single-lined. All measurements shown are phase-folded according to the orbital period and ephemeris obtained from lightcurve modelling.}
\label{fig:RVs}
\end{center}
\end{figure*}

We acquired phase-resolved spectra for all objects in the sample; however, at this time, we were only able to extract reliable radial velocity solutions for the single-lined systems. For these systems, we measured velocities by fitting Voigt profiles to the Balmer series of absorption lines and to helium absorption lines, when present. Single-lined spectroscopic binaries show a radial velocity from only one of their two components, whereas double-lined systems contain contributions from both. Double-lined systems are in principle more valuable, as one can precisely constrain the mass ratio of the system, and if combined with an inclination constraint from lightcurve modelling, solve for both of the dynamical masses in the system using the binary mass functions
\begin{equation}
    \label{eq:BMF}
    \frac{M_A^3{\sin}^3(i)}{(M_A+M_B)^2}=\frac{P_bK_B^3}{2\pi G}~,
\end{equation}
and 
\begin{equation}
    \label{eq:BMF2}
    \frac{M_B^3{\sin}^3(i)}{(M_B+M_A)^2}=\frac{P_bK_A^3}{2\pi G}~.
\end{equation}where $K_A$ and $K_B$ are the radial velocity semi-amplitudes of the two components, $i$ is the inclination of the system, and $P_b$ is the orbital period.
These expressions arise from Newtonian mechanics, and thus, the physics involved is straightforward, allowing for a more robust estimate of masses than those which invoke more complicated physical models, such as WD mass-radius relations, evolutionary tracks, and atmospheric models. However, in practice, fitting double-lined WDs is challenging when the components are both faint and in a short orbital period binary. One reason is that the absorption lines in WDs are intrinsically broad, such that their overlap is significant even at maximum radial velocity shift (the can be several hundred $\rm \AA$ across, whereas the Doppler shifts correspond to on the order of 10 $\rm \AA$). WDs do have non local thermodynamic equilibrium (NLTE) cores in their hydrogen absorption lines with a narrower full width half maximum (FWHM) allowing for more precise radial velocity measurements; however, resolving these line cores require moderate resolution spectra at good SNR. In binary systems with short orbital periods, exposures must be short to avoid significant orbital smearing, and this results in too few photons per exposure to reliably use medium resolution spectra without suffering from readout noise. Because of these technical challenges, the work to measure the dynamical masses of the double-lined systems of our sample (e.g. ZTF J0722$-$1839) is ongoing. See Figure \ref{fig:RVs} for an illustration of well-measured radial velocities in our sample (for the single-lined systems), and Table \ref{tab:measured_quantities} for the measured radial velocity semi-amplitudes of objects in our sample.

\begin{deluxetable*}{ccccccc}[htbp]
\tablenum{3}
\tablecaption{Orbital Characteristics \label{tab:measured_quantities}}
\tablewidth{0pt}
\tablehead{\colhead{Quantity:} & \colhead{$ K_{A}$} & \colhead{$ K_{B}$} & \colhead{$ i $} & \colhead{$ T_{0}$}  & \colhead{$ P_{b}$} \\
 & \colhead{($\rm km\,s^{-1}$)} & \colhead{($\rm km\,s^{-1}$)} & \colhead{($\rm deg$)} & \colhead{($\rm MBJD_{TDB}$)}  & \colhead{($\rm s$)} 
}

\startdata
ZTF J1539+5027$^1$ & $292^{+254}_{-583}$   & $961^{+178}_{-139}$ &$84.15^{+0.64}_{-0.57}$&$58305.1827886(12)$&$414.7915404(29)$&\\
ZTF J0538+1953 & & &$85.43^{+0.07}_{-0.09}$ & $58734.23560684(49)$&$866.60331(16)$ &\\
ZTF J1905+3134 &   & &&&	 $1032.16441(62)$ \\
PTF J0533+0209$^2$ &  & $618.7^{+6.9}_{-6.9}$ &$72.8^{+0.8}_{-1.4}$&$58144.596848(46)$&$1233.9729(17)$&\\
ZTF J2029+1534 &   &&$86.64^{+0.70}_{-0.4}$&$58757.255378(55)$&$1252.056499(41)$& \\
ZTF J0722$-$1839 &  & &$89.66^{+0.22}_{-0.22}$ & $58874.1823868(11)$&$1422.548655(71)$\\
ZTF J1749+0924 &   &  & $85.45^{+1.40}_{-1.15}$ & $58634.41086(2)$ &$1586.03389(44)$\\
ZTF J2228+4949 &   &  & &&$1713.58282(25)$\\
ZTF J1946+3203 &  & $284.8^{+4.8}_{-4.8}$& $77.08^{+1.6}_{-1.2}$ &$58836.38825(13)$&$2013.82141(75)$\\
ZTF J0643+0318 &   & $432.5^{+12.0}_{-12.0}$ & &&$2214.8058(11)$\\
ZTF J0640+1738 &   & $315.9^{+8.0}_{-8.0}$  & $65.3^{+5.1}_{-5.1}$ & $58836.38825(13)$&$2236.0160(16)$&\\
ZTF J2130+4420$^3$ &  &  $418.5^{+2.5}_{-2.5}$ & $86.4^{+1.0}_{-1.0}$ &$58672.1808578(1)$&$2360.4062(14)$&\\
ZTF J1901+5309 &   & & $87.28^{+0.52}_{-0.50}$ &$58703.3738042(31)$&$2436.10817(93)$&\\
ZTF J2320+3750 &   & $466.0^{+9.0}_{-9.0}$ & $84.5^{+2.7}_{-3.2}$ &$58769.28488(28)$&$3314.7998(35)$&\\
ZTF J2055+4651$^4$  & &$404.0^{+11.0}_{-11.0}$&$83.3^{+0.5}_{-0.5}$&$58731.944425(1)$&$3380.8701(29)$&\\
\enddata
\tablecomments{Measured orbital parameters for all of our systems, including the radial velocity semi-amplitudes of the components, $K_{A}$ and $K_{B}$, the orbital inclination, $i$, the time of superior conjunction, $T_{0}$, and the orbital period, $P_{b}$. All quantities were determined using the analyses described in Section 3, with the exception of ZTF J1539+5027, ZTF J0533+0209, ZTF J2130+4420 and ZTF J2055+4651, whose parameters were taken from previous publications. \tablerefs{\citet{Burdge2019a}$^1$, \citet{Burdge2019b}$^2$, \citet{Kupfer2020a}$^3$, \citet{Kupfer2020b}$^4$}}
\end{deluxetable*}

\subsection{Spectroscopic modelling}

In addition to being used to measure radial velocities, spectra can be used to estimate the temperature, surface gravity, and atmospheric abundances of stars. For single-lined systems, we fit atmospheric models to the entire spectrum in order to estimate the temperature of the dominant component, and supply this temperature to our lightcurve modelling. We perform an iterative process for double-lined systems, due to the spectra being a blend of two WDs (one must account for this, because cool WDs often have deep absorption lines, and thus, even if they contribute only a small fraction of the overall luminosity, they can significantly impact line profiles). First, we fit a single WD model to the spectrum to estimate the temperature of the dominant component. We then proceed to model the lightcurve while fixing the temperature of the hotter component to this value. Using this model, we can estimate both a surface brightness ratio, and ratio of component radii. We use this information to construct a composite model of two WD atmospheres, and we fit this composite model to the spectra to arrive at the final temperature estimates. For the DA WD systems we use the NLTE DA models described in \citet{Tremblay2011} with Stark broadening from \citet{Tremblay2009}. For the coolest and lowest surface gravity DA in our sample, ZTF J2320+3750, we use models described in \citet{Gianninas2014,Tremblay2015}. For modelling lower surface gravity ($\log(g)$) mixed H/He atmosphere objects which could be either hot subdwarfs or young He WDs, we used the model spectra described in \citet{Stroeer2007}. Please note, that when we refer to He WDs throughout this publication, it indicates helium core composition, not atmospheric composition. We did not factor spectroscopically measured surface gravity into our final parameter estimation, except for in the case of ZTF J2320+3750, in which the He WD component was poorly constrained by mass-radius relations alone due to the significant temperature sensitivity of these relations at such low mass.

\subsection{Orbital Evolution}

The orbital evolution of short orbital period binary systems is dominated by energy loss due to the emission of gravitational radiation, but can also be strongly influenced by the presence of mass transfer in accreting systems, especially ones with large mass ratios such as AM CVns \citep{Postnov2014}. Measuring orbital evolution in detached systems is particularly valuable, as general relativity predicts that the orbital frequency derivative, $\dot{f}_{GW}$, is directly related to the gravitational-wave frequency $f_{GW}=\frac{2}{P_b}$ and the chirp mass of the system,  $\mathcal{M}=\frac{(M_AM_B)^{\frac{3}{5}}}{(M_A+M_B)^{\frac{1}{5}}}$, as given by

\begin{equation}\label{eq:Decay}
   \dot{f}_{GW}=\frac{96}{5}\pi^{\frac{8}{3}} \left(\frac{GM_c}{c^3}\right)^\frac{5}{3}f_{GW}^\frac{11}{3},
\end{equation} \citep{Taylor1989}.

Conveniently, the strain of gravitational radiation emitted from such a binary depends only on the orbital frequency, chirp mass, and distance to the object, and thus the gravitational wave strain \emph{LISA} will measure is given by
\begin{equation}
    \label{eq:strain}
   S=\frac{2(G\mathcal{M})^{5/3}(\pi f_{GW})^{2/3}}{c^{4}D},
\end{equation} 

Multiplying this quantity by $\sqrt{f_{GW} T_{\rm obs}}$ gives the characteristic strain, where $T_{\rm obs}$ is the total time \emph{LISA} has operated \citep{Thorne1987,Moore2015}. For the characteristic strains computed in this work, $T_{\rm obs}$ is taken to be 4 years.

By constraining the chirp mass and orbital frequency from optical data, a \emph{LISA} gravitational-wave signal can be used to infer a distance to the system, which will be particularly useful since most of the objects being discovered are too faint for reliable Gaia astrometric solutions. For over-constrained systems where the distance is well determined via a measurement like the parallax, the gravitational-wave strain could be compared with that predicted based on an orbital decay rate due to purely general relativity, to estimate a deviation in the decay rate from tidal dissipation. Thus far, orbital decay has been detected and reported for two objects in the sample: ZTF J1539+5027 \citep{Burdge2019a}, and PTF J0533+0209 \citep{Burdge2019b}. The precisely measured decay rates in these two systems were enabled by regular photometric monitoring combined with well-sampled archival Palomar Transient Factory data collected a decade ago. We have not reached this threshold for any other systems, but continue to monitor all systems for orbital frequency evolution. The uncertainties of the ephemerides reported in Table \ref{tab:measured_quantities} give a characteristic estimate of the precision with which each system can be timed, and this can be compared to the predicted phase-shift in timing residuals given the expected decay rate of the system (see \citet{Burdge2019a} for more details) in order to estimate how long it will take to measure the orbital evolution precisely. The period derivative should be easily detectable in all systems before \emph{LISA} begins to operate if their orbital evolution is dominated by the emission of gravitational waves.

\subsection{Distance Estimates}

We report our estimated distances to the systems in Table \ref{tab:distances}. Estimating the distance to systems in our sample is an important element in this work as they are crucial in computing accurate gravitational wave amplitudes for the sources as measured from the solar system. Most of the systems in our sample have temperature estimates, as well as radius estimates; this allows us to compute distances directly based on these values, and synthetic spectra for WDs. We use system parameters inferred from a combination of lightcurve modelling and spectroscopic fitting to generate a composite synthetic spectrum for each system, and compute a synthetic apparent magnitude from these spectra. We chose to compute Pan-STARRS1 $g$-band apparent magnitudes, and comparing to the measured value. We use Pan-STARRS1 $g$ due to its availability for all objects and its short wavelength compared to other available passbands, ensuring minimal potential contamination from blended background sources in dense fields (which on average, are likely to contribute more to redder bands). These spectroscopic distance estimates are used in computing the \emph{LISA} SNR. We estimate uncertainties on these distance estimates by setting up a Monte Carlo simulation in which we sample over distance and the temperatures and radii of both WDs (with priors set by the lightcurve+spectroscopic analysis). During each iteration, we use the two temperatures and radii to generate a synthetic spectrum, and automatically query the reddening catalog described in \citet{Green2019} at the coordinates of the source using the distance sample from that iteration to estimate the redenning along the line of sight, and use this to apply a reddening correction to the synthetic photometry we generate. We then compare this synthetic flux to the measured Pan-STARRS1 $g$, and compute the likelihood function using the uncertainty reported by Pan-STARRS1. The spectroscopic distances reported in Table \ref{tab:distances} were computed using the posterior distribution of this analysis.

It is worth noting that \emph{Gaia} has revolutionized astronomy by providing the capability to estimate distances to a large number of stars in the Galaxy via the measurement of parallax. However, for faint populations of sources with uncertain parallaxes, there are subtleties to inferring distance from the parallax, $\bar{\omega}$. Nonetheless, we wish to emphasize that when available, a precisely measured \emph{Gaia} parallax is incredibly valuable, as this method of estimating distances arises from a simple geometric effect, whereas other techniques of estimating distance invoke more complicated model dependence. Here, we compare two approaches for estimating distances the sample of ZTF short period binaries using \emph{Gaia} parallaxes, though the technique we ultimately use in this paper is the spectroscopic distance estimate described above.

The two parallax-based distance estimates presented in Table \ref{tab:distances} use an exponentially decreasing space density prior as described in \citet{Bailer-Jones2018}. An important element of this prior is that it invokes a characteristic length scale (note, this is not a scale height). The distances reported by \citet{Bailer-Jones2018} couple this prior with length scales derived from a model of the galaxy. However, this model has its shortfalls when considering a specific population of objects such as WDs, as it is purely geometric and intended to be relatively independent of the properties of stars, and thus is not an optimal choice for characterizing a sub-population of stars where additional information is present. Notably, as illustrated in Table \ref{tab:distances}, the Bailer-Jones technique tends to yield large distances at low Galactic latitudes, where the characteristic length scale is significantly larger than at higher Galactic latitude. Because WDs are foreground objects and this length scale is estimated based on all the stars in a region of the sky, this method likely overestimates distances in the Galactic plane for WD populations. The alternative method, adopted by \citet{Kupfer2018}, invokes the same exponentially decreasing space density prior, but assumes a $400\,\rm pc$ characteristic length scale for all sources based on WD population synthesis work, rather than a length scale dependent on sky position. We report estimates based on a modified version of this method, with the same maximum likelihood measurement, but instead use a 68 percent credible interval derived from the highest posterior density of our distribution to estimate the error bars.

\begin{deluxetable*}{ccccccccc}[htbp]
\tablenum{4}
\tablecaption{Distance and Reddening Estimates \label{tab:distances}}
\tablewidth{0pt}
\tablehead{\colhead{Technique:} & \colhead{Spectroscopic} & \colhead{Bailer-Jones}  & \colhead{400 pc Length Scale}   & \colhead{E(g-r)}  &\colhead{\emph{Gaia} $\bar \omega$}&\colhead{\emph{Gaia} $\mu$ RA}&\colhead{\emph{Gaia} $\mu$ Dec} \\
 & \colhead{($\rm kpc$)} & \colhead{($\rm kpc$)}   & \colhead{($\rm kpc$)}&\colhead{($\rm m_{AB}$)} &\colhead{($\rm mas$)} &\colhead{($\rm mas\,yr^{-1}$)} &\colhead{($\rm mas\,yr^{-1}$)} }

\startdata
ZTF J1539+5027 &$2.34^{+0.14}_{-0.14}$& $1.39^{+0.87}_{-0.53}$  & $1.25^{+0.72}_{-0.45}$  & &$-0.11\pm0.79$&$-3.41\pm1.53$&$-3.82\pm1.58$ \\
ZTF J0538+1953&$0.684^{+0.018}_{-0.014}$ &  $1.07^{+1.48}_{-0.45}$ & $0.84^{+0.46}_{-0.26}$  &$0.25^{+0.02}_{-0.02}$ &$1.15\pm0.48$&$-2.22\pm0.54$&$-5.67\pm0.45$\\
ZTF J1905+3134 & $3.8-9.5$  & &&  \\
PTF J0533+0209 &$1.74^{+0.14}_{-0.14}$ &  $1.54^{+1.06}_{-0.58}$ & $1.23^{+0.63}_{-0.40}$  &$0.11^{+0.02}_{-0.02}$& $0.47\pm0.48$&$1.43\pm0.74$&$2.56\pm0.91$\\
ZTF J2029+1534 & $2.02^{+0.15}_{-0.15}$&  $2.39^{+1.89}_{-1.17}$  & $1.17^{+0.74}_{-0.47}$  &$0.08^{+0.03}_{-0.01}$ &$-1.43\pm1.43$&$-5.94\pm1.51$&$-9.92\pm1.93$\\
ZTF J0722$-$1839 &$0.928^{+0.023}_{-0.031}$ & $2.51^{+2.28}_{-1.10}$ & $1.41^{+0.61}_{-0.42}$  &$0.14^{+0.03}_{-0.01}$& $0.41\pm0.35$ & $-5.83\pm0.52$ &$4.18\pm0.65$\\
ZTF J1749+0924 &$1.55^{+0.20}_{-0.18}$& $3.34^{+2.55}_{-1.58}$ & $1.34^{+0.76}_{-0.50}$  &$0.13^{+0.03}_{-0.01}$& $-2.68\pm1.36$&$-6.50\pm2.78$&$6.90\pm3.74$\\
ZTF J2228+4949 &$0.72-1.77$ & $1.79^{+1.48}_{-0.67}$  & $1.31^{+0.54}_{-0.36}$  &$0.14^{+0.02}_{-0.02}$&$0.57\pm0.31$&$-3.83\pm0.56$&$-2.56\pm0.53$ \\
ZTF J1946+3203 &$2.38^{+0.18}_{-0.20}$&  $2.18^{+4.12}_{-1.08}$ & $1.15^{+0.51}_{-0.33}$  &$0.14^{+0.02}_{-0.04}$&$0.72\pm0.35$&$-1.86\pm0.46$&$-4.27\pm0.49$ \\
ZTF J0643+0318 & &  $2.13^{+1.59}_{-0.74}$ & $1.55^{+0.51}_{-0.39}$ &$0.45^{+0.02}_{-0.05}$&$0.47\pm0.22$&$0.71\pm0.40$&$1.50\pm0.41$ \\
ZTF J0640+1738 &$5.9^{+1.3}_{-1.2}$ &  $1.90^{+1.61}_{-0.80}$ & $1.24^{+0.61}_{-0.39}$  &$0.12^{+0.02}_{-0.02}$ & $0.51\pm0.44$& $-0.36\pm1.16$& $1.10\pm1.33$\\
ZTF J2130+4420 & $1.209^{+0.046}_{-0.047}$ & $1.116^{+0.043}_{-0.040}$ & $1.16^{+0.10}_{-0.08}$   &$0.18^{+0.02}_{-0.02}$&$0.833\pm0.031$&$0.009\pm0.031$&$-1.682\pm0.0048$ \\
ZTF J1901+5309 & $0.831^{+0.050}_{-0.047}$ & $0.89^{+0.11}_{-0.09}$  & $0.88^{+0.08}_{-0.10}$  &$0.05^{+0.02}_{-0.02}$& $1.11\pm0.12$&$2.94\pm0.25$&$4.75\pm0.26$ \\
ZTF J2320+3750 & $2.51^{+0.22}_{-0.22}$  & $1.29^{+0.79}_{-0.42}$  & $1.16^{+0.54}_{-0.35}$  &$0.15^{+0.01}_{-0.01}$&$0.68\pm0.39$&$13.00\pm0.49$&$11.74\pm0.43$ \\
ZTF J2055+4651 & $2.57^{+0.12}_{-0.13}$ & $2.17^{+0.63}_{-0.41}$  & $1.94^{+0.30}_{-0.36}$  &$0.53^{+0.02}_{-0.01}$ &$0.432\pm0.098$&$-3.33\pm0.17$&$-5.16\pm0.17$ \\
\enddata
\tablecomments{Comparison of three methods for estimating distances to the sample of binaries, including a spectroscopic distance, and two parallax based distances. Note that the Bailer-Jones \citep{Bailer-Jones2018} and 400 pc length scale \citep{Kupfer2018} techniques depend on \emph{Gaia} astrometry, while the spectroscopic distances are determined from the measured radii and temperatures of the objects. We also estimate the reddening, E(g-r), for each system \citep{Green2019}, and present the measured \emph{Gaia} parallax, $\bar \omega$, and proper motions , $\mu$ RA and $\mu$ Dec, for each system. }
\end{deluxetable*}

\subsection{Combined Analysis and Observed Characteristics}

We conduct a combined analysis of the lightcurve modelling of each system and modelling of the phase-resolved spectroscopy in order to estimate the properties of the systems reported in Table \ref{tab:derived_quantities}. We conducted the combined analysis using the nested sampling package MULTINEST \citep{Feroz2009}. When sampling, we combined modelling the lightcurve with priors on the temperature of the hotter component in each binary based on a spectroscopic fit, and for the single-lined systems, an additional constraint based on the measured radial-velocity semi-amplitude. For single-lined systems, we used the velocity semi-amplitude constraint plus constraints on the mass ratio from ellipsoidal modulation to infer parameters for the system. For double-lined systems, we instead used the mass-radius relations described in \citet{Soares2017} to constrain the system parameters, allowing us to infer masses from radii estimated using the lightcurve. The single-lined, eclipsing, and ellipsoidal variable ZTF J1946+3203 is an exception, as in this system we inferred masses from a combination of the radial velocity semi-amplitude, and well-constrained inclination and mass ratio inferred from the lightcurve, as well as radii measurable by the eclipses. Thus, this system serves as an example of one with parameters that can be used in testing mass-radius relations. Note, that we did not model the two AM CVn systems, ZTF J1905+3134 and ZTF J2228+4949, and that modelling of the mass transferring system ZTF J0643+0318 is still underway.

\begin{deluxetable*}{ccccccc}[htbp]
\tablenum{5}
\tablecaption{Physical System Parameters \label{tab:derived_quantities}}
\tablewidth{0pt}
\tablehead{\colhead{Quantity:} & \colhead{$M_{A}$} & \colhead{$M_{B}$} & \colhead{$R_{A}$} & \colhead{$R_{B}$}  & \colhead{$T_{A}$} & \colhead{$T_{B}$}  \\
\colhead{Units:} & \colhead{($M_{\odot}$)} & \colhead{($M_{\odot}$)} & \colhead{($R_{\odot}$)} & \colhead{($R_{\odot}$)}  & \colhead{($\rm kK$)} & \colhead{($\rm kK$)}  
}

\startdata
ZTF J1539+5027$^1$ &  $0.610^{+0.017}_{-0.022}$ & $0.210^{+0.015}_{-0.015}$ &$0.01562^{+0.00038}_{-0.00038}$&$0.03140^{+0.00054}_{-0.00052}$&$48.9^{+0.9}_{-0.9}$&$<10$\\
ZTF J0538+1953 & $0.45^{+0.05}_{-0.05}$ & $0.32^{+0.03}_{-0.03}$ &$0.02069^{+0.00028}_{-0.00034}$& $0.02319^{+0.00032}_{-0.00026}$&$26.45^{+0.725}_{-0.725}$&$12.8^{+0.2}_{-0.2}$\\
PTF J0533+0209$^2$ &  $0.652^{+0.037}_{-0.040}$ & $0.167^{+0.030}_{-0.030}$  &&$0.057^{+0.004}_{-0.004}$&&$20^{+0.8}_{-0.8}$\\
ZTF J2029+1534 & $0.32^{+0.04}_{-0.04}$ &$0.30^{+0.04}_{-0.04}$&$0.029^{+0.002}_{-0.003}$&$0.028^{+0.003}_{-0.003}$&$18.25^{+0.25}_{-0.25}$&$15.3^{+0.3}_{-0.3}$ \\
ZTF J0722$-$1839 & $0.38^{+0.04}_{-0.04}$ & $0.33^{+0.03}_{-0.03}$ &$0.0224^{+0.0004}_{-0.0002}$ &$0.0249^{+0.0001}_{-0.0003}$ &$19.9^{+0.15}_{-0.15}$&$16.8^{+0.15}_{-0.15}$\\
ZTF J1749+0924 &  $0.40^{+0.07}_{-0.05}$ & $0.28^{+0.05}_{-0.04}$ & $0.022^{+0.003}_{-0.004}$ &$0.025^{+0.004}_{-0.004}$&$20.4^{+0.2}_{-0.2}$&$12.0^{+0.6}_{-0.6}$\\
ZTF J1946+3203 & $0.272^{+0.046}_{-0.043}$  & $0.307^{+0.097}_{-0.085}$ & $0.0299^{+0.0049}_{-0.0045}$ & $0.111^{+0.012}_{-0.013}$ &$11.5^{+2.3}_{-4.6}$ &$28.0^{+1.7}_{-1.7}$\\
ZTF J0640+1738 &   $0.39^{+0.12}_{-0.089}$  & $0.325^{+0.30}_{-0.15}$  &  &$0.152^{+0.040}_{-0.032}$&$10.2^{+8.3}_{-6.8}$&$31.5^{+5.0}_{-5.0}$\\
ZTF J2130+4420$^3$ &  $0.545^{+0.020}_{-0.020}$ &$0.337^{+0.015}_{-0.015}$   & &$0.125^{+0.005}_{-0.005}$&&$42.4^{+3.0}_{-3.0}$\\
ZTF J1901+5309 &  $0.36^{+0.04}_{-0.04}$  & $0.36^{+0.05}_{-0.05}$ &$0.029^{+0.001}_{-0.002}$ &$0.022^{+0.003}_{-0.002}$ &$26.0^{+0.2}_{-0.2}$&$16.5^{+2.0}_{-2.0}$ \\
ZTF J2320+3750 & $0.69^{+0.03}_{-0.03}$  & $0.20^{+0.01}_{-0.01}$ &  &$0.152^{+0.015}_{-0.017}$&&$9.2^{+0.2}_{-0.2}$\\
ZTF J2055+4651$^4$  & $0.68^{+0.05}_{-0.05}$ &$0.41^{+0.04}_{-0.04}$&$0.0148^{+0.002}_{-0.002}$&$0.17^{+0.01}_{-0.01}$&$63.0^{+1.0}_{-1.0}$&$33.7^{+1.0}_{-1.0}$\\
\enddata
\tablecomments{Derived physical parameters for the components in our systems, including the masses of the two components, $M_{A}$ and $M_{B}$, the radii of the components, $R_{A}$ and $R_{B}$, and the effective surface temperatures of the components, $T_{A}$ and $T_{B}$. We have excluded systems with significant mass transfer contributing to the luminosity: ZTF J0643+0318, ZTF J1905+3134, and ZTF J2228+4949. All were determined using the analyses described in section 3, with the exception of ZTF J1539+5027, PTF J0533+0209, ZTF J2130+4420 and ZTF J2055+4651, whose parameters were taken from previous publications.\tablerefs{\citet{Burdge2019a}$^1$, \citet{Burdge2019b}$^2$, \citet{Kupfer2020a}$^3$, \citet{Kupfer2020b}$^4$}}
\end{deluxetable*}

\subsection{\emph{LISA} Signal to Noise}

In order to estimate the \emph{LISA} gravitational wave SNR, we adopt the same formalism as outlined \citet{Burdge2019a} and \citet{Burdge2019b} (which was based on the formalism outlined in \citet{Korol2017}), correcting a factor of 2 error which was present in the previous estimates, resulting in overall lower signal-to-noise (SNR) estimates. We use the updated sensitivity curve presented in \citet{Robson2019}.

For a more detailed discussion of \emph{LISA} SNR estimates, and their subtleties, please see the appendix.

\section{Discussion of individual systems} \label{sec:systems}

\subsection{ZTF J1539+5027}

ZTF J1539+5027 is a deeply eclipsing DWD binary system with an orbital period of just $\approx 6.91 \,\rm min$ \citep{Burdge2019a} consisting of a cool He WD orbiting a heated CO WD. This is one of two systems in our sample with confirmed orbital decay, and thus, a precise measurement of its chirp mass (assuming the decay is due to energy loss to gravitational radiation). There is ongoing work to more precisely characterize this system using observations obtained with the high speed photometer HiPERCAM (see the first panel in Figure \ref{fig:FollowUpPhot}) and the COS spectrograph on the Hubble Space Telescope. The system is among the highest SNR \emph{LISA}-detectable gravitational-wave sources known, and presents a curious evolutionary puzzle, as the CO WD in the system has been heated to $\approx 50,000 \, \rm K$, but there are no observational indications of active mass transfer (see \citet{Burdge2019a} for further details).

\subsection{ZTF J0538+1953}

ZTF J0538+1953 is a $14.44$-min orbital period eclipsing detached DWD system. The hotter WD is $\approx26,000\,\rm K$, and the cooler companion $\approx13,000\,\rm K$. The two components are of comparable radii (with the cooler being slightly larger than its hot companion, indicating that it is likely the less massive component). The lightcurve exhibits a moderate degree of flux reprocessed by the heated face of the cooler WD in the system (causing an 8 percent peak-to-peak amplitude photometric modulation on the orbital period), though this effect is much smaller than that observed in ZTF J1539+5027. After conducting lightcurve modelling, we found a best fit for the mass ratio $q=\frac{M_B}{M_A}=5.01\pm0.55$, where $M_B$ corresponds to the mass of the cooler, larger secondary. This is highly unlikely because this physically conflicts with WD mass-radius relations (which were used to infer the final reported masses).

Upon further inspection of the residuals of the model, we see that even with the best-fit model, there is a residual signal with a few percent amplitude at twice the orbital frequency, phase-shifted with respect to the expected phase of ellipsoidal modulation. This signal could arise as a result of incomplete modelling of the reprocessed radiation on the heated face of the secondary, or potentially, be indicative of a dynamically excited tidal pulsation, as is expected to occur at twice the orbital frequency in circular orbit systems. Further follow-up should allow us to understand this signal, but for the purposes of this work we have omitted quoting a mass ratio of the object based on the lightcurve, as it is difficult to disentangle this signal from any weak ellipsoidal modulation that may be present. 

ZTF J0538+1953 is perhaps best compared to SDSS J0651+2844, the $12.75$-min orbital period detached binary discovered by \citet{Brown2011} and further characterized in \citet{Hermes2012}. ZTF J0538+1953's components have measured $R/a$ values of $\approx 0.11$ for the hot $\approx26,000\,\rm K$ and $\approx0.13$ for the cooler $\approx13,000\,\rm K$ component, whereas the components of SDSS J0651+2844 have $R/a$ values of $\approx0.12$ for the hotter object and $\approx0.045$ for the cooler, more compact companion. 

Invoking mass-radius relations in combination with the lightcurve modelling (which make use of the radii inferred from eclipses to determine masses), we find that the cooler component in ZTF J0538+1953 is consistent with a $0.32^{+0.03}_{-0.03}\,\rm M_\odot$ object, suggesting it is a He WD. The hotter WD in the system has a mass consistent with a $0.45^{+0.05}_{-0.05}\, \rm M_\odot\ $ object, which lies in a regime where the WD could be a He WD, CO WD, or hybrid WD \citep{Perets2019}.

\subsection{ZTF J1905+3134}

ZTF J1905+3134 is an AM CVn system exhibiting a photometric modulation at $\approx17.2$ minutes, and strong double-peaked \ion{He}{2} emission, similar to that seen in SDSS J1351$-$0643 \citep{Green2018}. Shorter orbital period AM CVn systems like SDSS J1351$-$0643 and ZTF J1905+3134 do not normally undergo outbursts like their longer period counterparts, and instead remain in a constant high state \citep{Ramsay2018}, and ZTF J1905+3134's strong \ion{He}{2} emission is characteristic of shorter period systems in the high state. The photometric modulation in this system could either originate from the orbital period or the disk precession period \citep{Green2018}; however, the deep, sharp temporal feature in the lightcurve (see Figure \ref{fig:FollowUpPhot}) suggests the system is eclipsing (with a possible additional eclipse of a hot spot shortly after the deeper eclipse), which would make this the shortest orbital period eclipsing AM CVn known. Further photometric follow-up of this faint system could constrain these features more precisely, but current follow-up (which consists of just a few orbital cycles) lacks the SNR necessary for detailed modelling. ZTF J1905+3134 is the only object in our sample which completely lacks a \emph{Gaia} astrometric solution, so we instead refer to \citet{Ramsay2018} in order to estimate the distance based on its luminosity and the period-luminosity relation for AM CVn systems. AM CVn systems around the period range of $17$ minutes exhibit absolute luminosity values in the range of $\rm M_G\approx 5-8$. Thus, using a uniform prior on the absolute luminosity reflecting this range, and comparing to the \emph{Gaia} apparent magnitude of $G=20.90\pm0.030$, we estimate that the object lies approximately $3.8-9.55\,\rm kpc$ from the Sun.

\subsection{PTF J0533+0209}

PTF J0533+0209 is a $\approx 20.6$-min orbital period detached binary system first described in \citet{Burdge2019b}. The system is unique due to its He dominated DBA atmosphere, which exhibits only traces of hydrogen~---a previously unseen characteristic among ELM WDs. The system's atmospheric composition raises question about formation channels because it requires fine-tuning of evolutionary models to produce an object with so little hydrogen (see \citet{Burdge2019b} for further details). Like ZTF J1539+5027, orbital decay has been detected in this system, enabling the measurement of a chirp mass assuming orbital evolution primarily due to general relativity. The system houses an unseen compact object, likely a CO WD. 

\subsection{ZTF J2029+1534}

ZTF J2029+1534 is a detached eclipsing DWD binary with an orbital period of $\approx 20.9$ minutes. With relatively large $R/a$ values of $\approx 0.15$ and $\approx 0.12$, and temperatures of $\approx 19,500\rm \,K$ and $\approx 17,000 \rm \,K$, respectively, this system likely consists of two similar temperature He WDs, with mass-radius relations giving estimated masses of $0.32^{+0.04}_{-0.04}\, M_\odot\ $ and $0.30^{+0.04}_{-0.04}\, M_\odot\ $ for the two components, respectively. The system is among the faintest in the sample, with a Pan-STARRS $g$-band apparent magnitude of $\approx 20.4$.

\subsection{ZTF J0722$-$1839}

\begin{figure}[htpb]
\begin{center}
\includegraphics[width=0.5\textwidth]{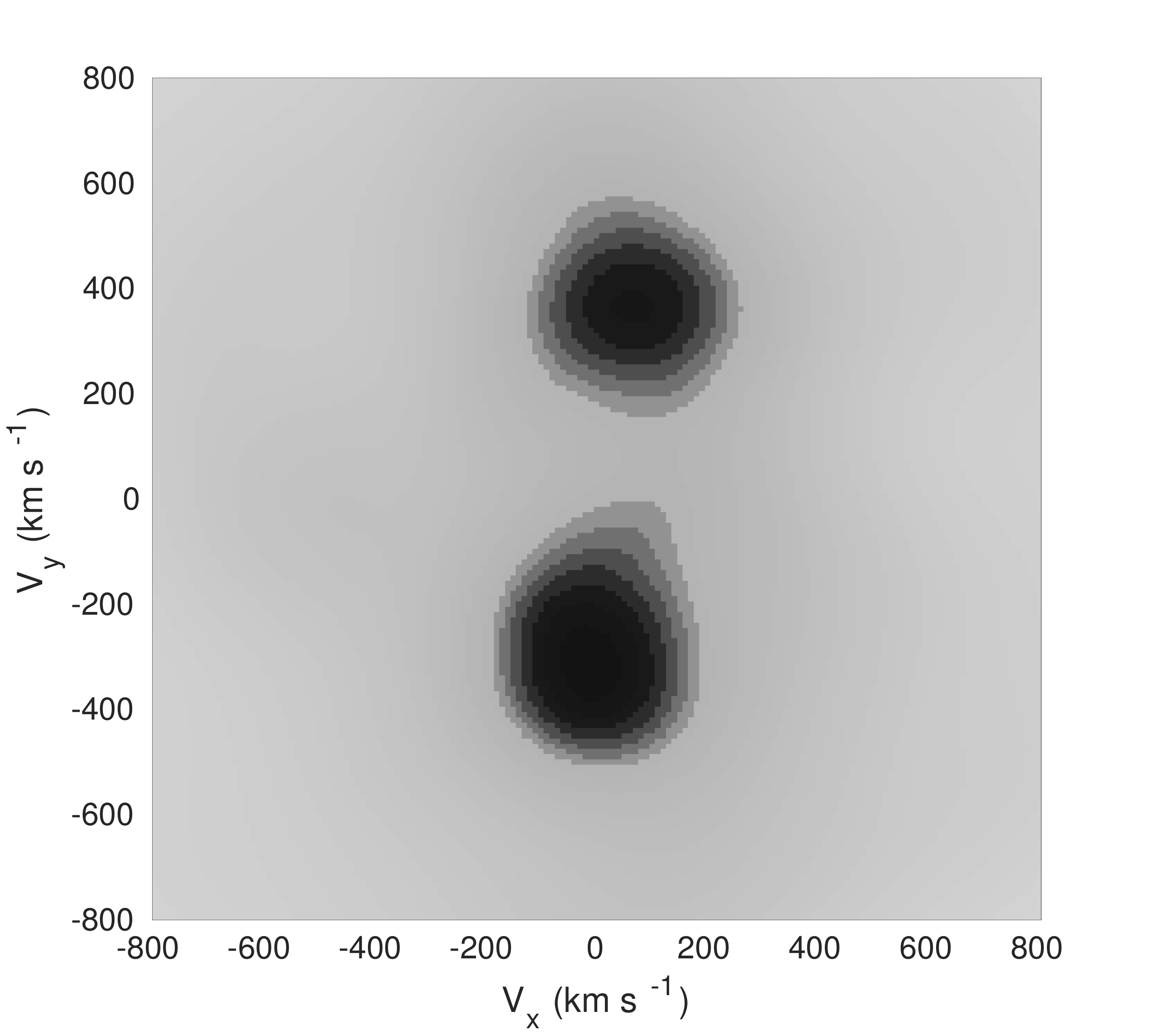}
\caption{Doppler tomogram \citep{Marsh2001} of the Balmer series of absorption lines in ZTF J0722$-$1839. The two dark features correspond to absorption line cores tracking each object, reflecting the system's double-lined nature. However, due to the shallow topographical nature of these features in the tomogram (as a result of overall low SNR spectra in each phase bin), the velocities of the two components are poorly constrained with the current set of spectra; however, further spectroscopic follow-up of this system is underway, as it is clearly a promising candidate for measuring dynamical masses, which would serve as a test of WD mass-radius relations.}
\label{fig:J0722}
\end{center}
\end{figure}

ZTF J0722$-$1839 is a detached eclipsing DWD binary with an orbital period of $\approx 23.70$ minutes. Remarkably, the primary and secondary eclipses in this system both result in an attenuation of nearly half of the system's flux (with the deeper eclipse attenuating by slightly over 50 percent of the flux, and the secondary eclipse over 40 percent of the flux). To exhibit two eclipses of this depth, the system must be fine-tuned. These eclipses indicate that the objects must be able to almost completely occult each other, indicating similar radii, and their nearly equal depths indicates nearly equal luminosity per unit surface area, implying similar temperatures for the two WDs. Finally, to achieve such deep eclipses, the system must be also be near edge on. Based on the system parameters reported in Table \ref{tab:measured_quantities}, we infer that this system is a double He WD system, with two He WDs of similar temperatures and radii. This may imply that the two objects have similar ages, requiring a formation mechanism capable of producing the two objects in rapid succession. The system is double-lined due to the comparable luminosity of both objects and similar line profiles due to the similarity in temperature and surface gravity (see Figure \ref{fig:J0722}). Further work is underway to measure precise dynamical masses in this system.

\subsection{ZTF J1749+0924}

ZTF J1749+0924 was the final discovery in the sample described in this paper, identified using the BLS algorithm. The system likely consists of a pair of eclipsing He WDs with an orbital period of approximately $26.4$ minutes. ZTF observed this system in a continuous cadence mode (taking exposures with an approximately $40\,\rm s$ cadence continuously for an hour and a half, on two separate occasions). On both occasions of continuously sampled observations by ZTF, the system exhibited multiple eclipses during the observation window (with each ``eclipse'' having only a single detection due to the short duty cycle of the feature). The system likely contains a pair of He WDs, with masses of  $0.40^{+0.07}_{-0.05}\,  M_\odot\ $ and  $0.28^{+0.05}_{-0.04}\,  M_\odot\ $.

\subsection{ZTF J2228+4949}

ZTF J2228+4949 is an AM CVn system whose spectrum exhibits double-peaked \ion{He}{2} emission characteristic of a hot accretion disk, much like that seen in ZTF J1905+3134. The system exhibits a photometric modulation at $\approx 28.56$ minutes; compared to other high state AM CVn systems \citep{Ramsay2018}, ZTF J2228+4949 this is an unusually long period. We propose that this system may be the longest period high state AM CVn system known, potentially due to an unusually high accretion rate at its orbital period. In any case, our survey, which targets systems displaying strong optical periodicity, is naturally biased towards systems in a persistent high state, as they are more likely to exhibit photometric modulation due to the presence of a superhump \citep{Green2018}, and also do not undergo outbursts, which can greatly impact the ability to search for periods. We estimate a spectroscopic distance of $0.72-1.77\,\rm kpc$, based on the typical absolute magnitude of AM CVns at these periods falling between $M_{\rm G}\approx 8-10$. Because ZTF J2228+4949 may have an unusually high accretion rate for its period, it is possible that the source is more distant than this spectroscopic estimate. A more precise distance estimate not based on the system's luminosity would be particularly valuable, as it would allow one to constrain whether the luminosity is indeed elevated compared to other sources at comparable orbital periods.

\subsection{ZTF J1946+3203}

\begin{figure*}[htpb]
\begin{center}
\includegraphics[width=1.0\textwidth]{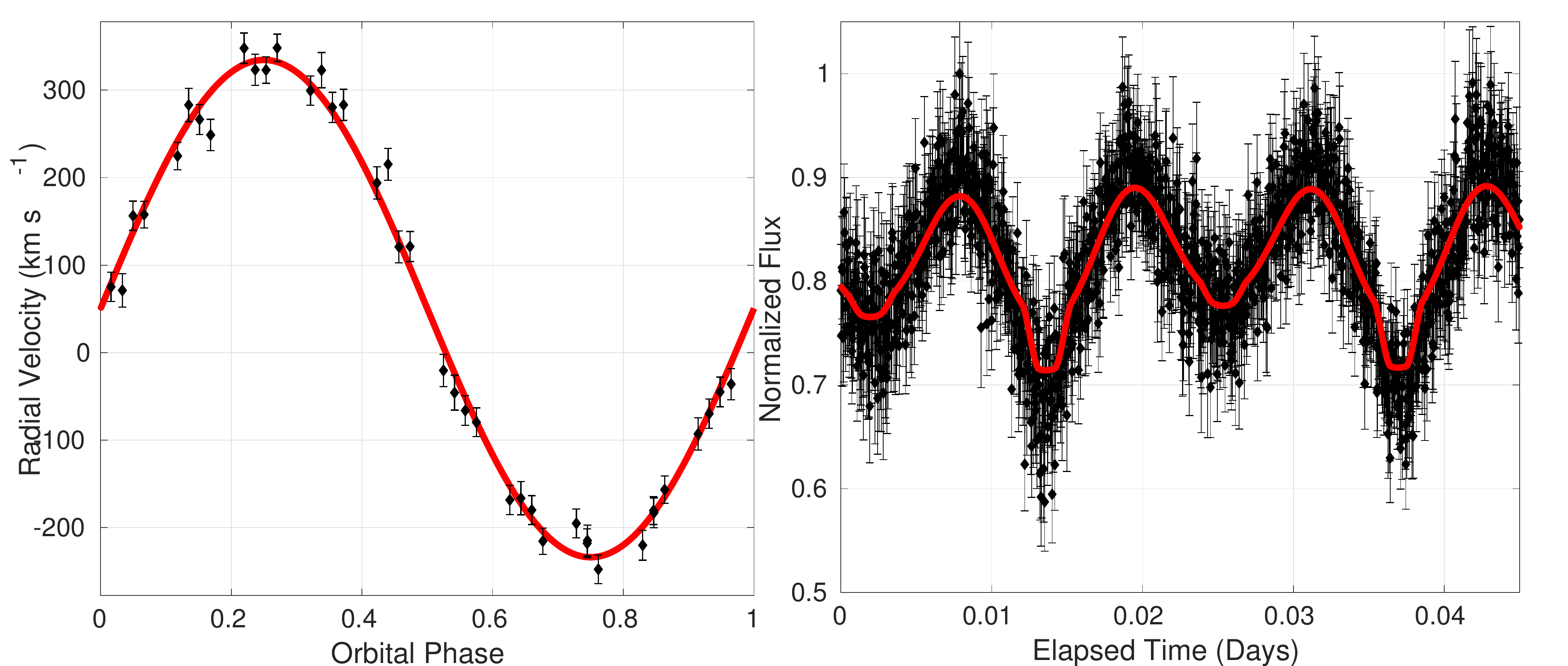}
\caption{Left Panel: Radial velocities of ZTF J1946+3203 as measured from phase-resolved spectra obtained with LRIS (black points). The smearing-corrected measured semi-amplitude for this system is $284.8\pm4.8 \rm \, km \, s^{-1}$, with the sinusoidal fit shown in red. Right panel: Best fit model (red) to the CHIMERA $g$-band lightcurve of the system (black points), which exhibits both strong ellipsoidal modulation, but also an eclipse. The presence of ellipsoidal modulation, eclipses, and a cleanly measured radial velocity semi-amplitude allowed us to measure both component masses and radii, which we have determined is likely a double He WD binary system. The data in this Figure is the same as shown in Figures 2 and 4, and is meant to illustrate an example model fit to the data used in parameter estimation.}
\label{fig:J1946}
\end{center}
\end{figure*}

ZTF J1946+3203 is a single-lined spectroscopic eclipsing binary which shows strong ellipsoidal variations (see Figure \ref{fig:J1946}). The more luminous component in this system has a surface temperature of approximately $28,000\pm2,000\,\rm K$ and a DAB atmosphere exhibiting both H and \ion{He}{1} lines, with a spectroscopically measured surface gravity of $\rm \log(g)=5.74\pm0.2$. A combined analysis of the lightcurve and radial velocities yields mass estimates of $0.272^{+0.046}_{-0.043}\,  M_\odot\ $ for the less luminous component (likely a He WD), and a mass of $0.307^{+0.097}_{-0.085}\,  M_\odot\ $ for the more luminous component (which could be consistent with either a hot He WD, or an sdB). Systems like ZTF J1946+3203 are particularly valuable, as they exhibit eclipses, ellipsoidal variations, and an unambiguous radial velocity semi-amplitude, strongly constraining component masses, temperatures, and radii.

\subsection{ZTF J0643+0318}

ZTF J0643+0318 is a single-lined spectroscopic eclipsing binary undergoing mass transfer, and consequently exhibiting a strong \ion{He}{2} emission feature. Further work is underway to characterize this system.  

\subsection{ZTF J0640+1738}

ZTF J0640+1738 is a single-lined spectroscopic binary exhibiting ellipsoidal modulation. The atmosphere of its more luminous component exhibits helium lines in addition to hydrogen, much like an sdB. Based on its large radius and high temperature, the system is likely a WD+sdB system. Because degenerate He WDs pass through the same location in color-luminosity space as sdBs, it is difficult to differentiate the two classes of objects. The parameters for this system are poorly constrained, due to the absence of a high quality follow-up lightcurve. Further follow-up is underway to better constrain this system's characteristics.

\subsection{ZTF J2130+4420}

ZTF J2130+4420 is a mass transferring WD+sdB system described in \citet{Kupfer2020a}. When the high-SNR HiPERCAM lightcurve of the system was modelled, the shape of the primary minimum in the lightcurve could only be reproduced with an accretion disk eclipsing the sdB in the system (which dominates the luminosity). The lightcurve of J2130+4420 and the similar system J2055+4651 are very distinctive, with minima of different depths, making them easy to identify as ellipsoidal/eclipsing variables and distinguish from other periodic variables in this period range.

\subsection{ZTF J1901+5309}

ZTF J1901+5309 is an eclipsing pair of He WDs described in detail in \citet{Coughlin2019}. While the core composition of the WDs is ambiguous due to the system being a double-lined spectroscopic binary, we have determined masses using mass-radius relations for the two components, and identified both as consistent with He WDs. 

\subsection{ZTF J2320+3750}

\begin{figure}[htpb]
\begin{center}
\includegraphics[width=0.5\textwidth]{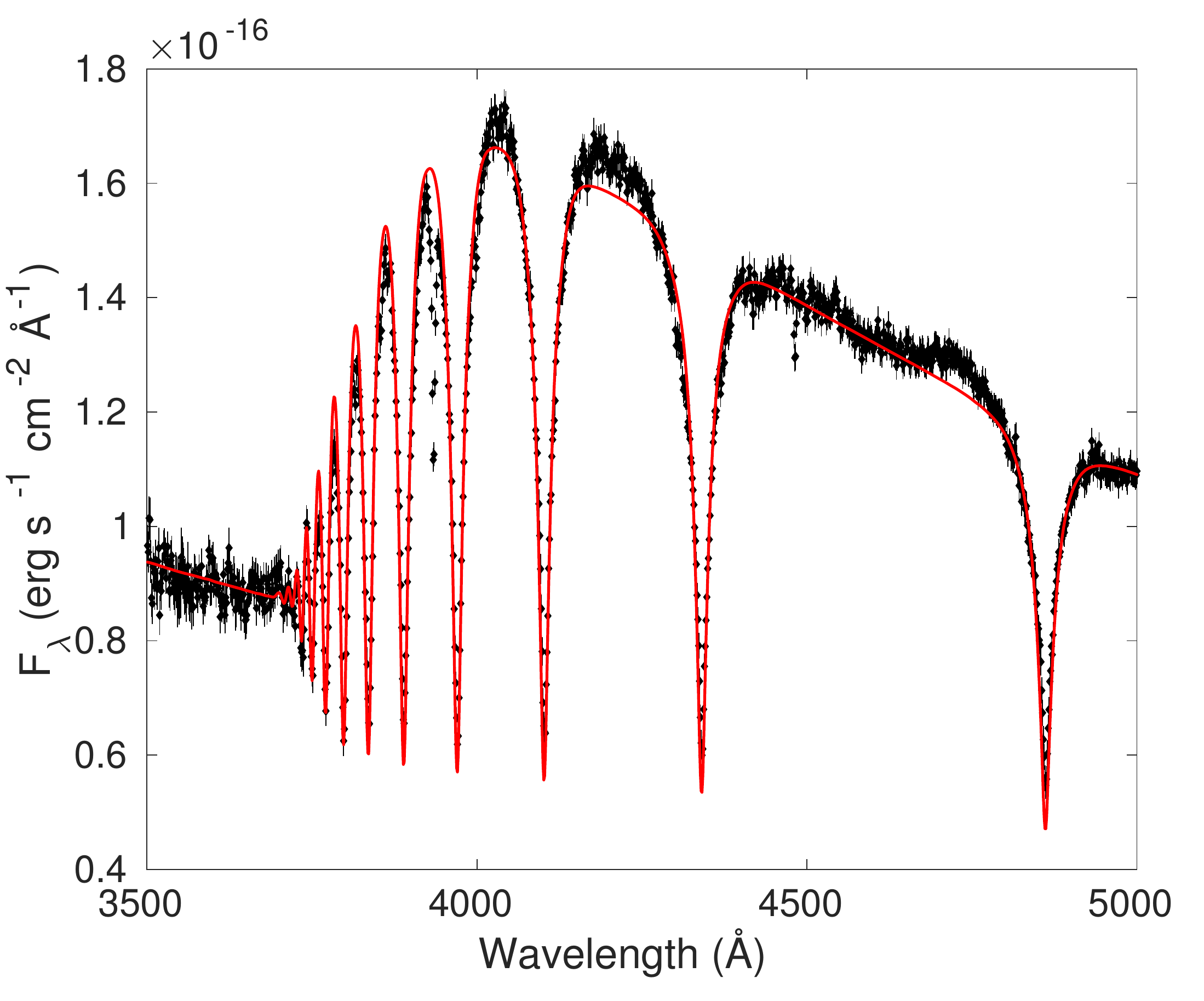}
\caption{The coadded and de-redenned LRIS spectrum of ZTF J2320+3750 (black points), fit with a spectroscopic model (red) to determine the surface temperature and gravity of the white dwarf.}
\label{fig:J2320s}
\end{center}
\end{figure}

ZTF J2320+3750 consists of a CO WD orbiting a He WD. The He WD exhibits a spectrum similar to other ELMs, including a Mg II metal line (ELMs often show enhanced photospheric abundances of metals, which has been attributed to a combination of rotational mixing and radiative levitation--see \citet{Kaplan2013,Hermes2014}). The spectrum of ZTF J2320+3750, with an example of the model used to fit it is illustrated in Figure \ref{fig:J2320s}. Notably, ZTF J2320+3750 has a large radial velocity semi-amplitude for its orbital period. Large-scale massively multiplexed spectroscopic surveys should be quite sensitive to detecting such systems with multiple observations.

\subsection{ZTF J2055+4651}

ZTF J2055+4651 is a mass-transferring and eclipsing WD+sdB system described in \citet{Kupfer2020b}. The system has a lightcurve which is incredibly similar to ZTF J2130+4420, with minima of very different depths, making it easily distinguishable from other variables such as $\delta$ Scuti pulsators. The high-SNR HiPERCAM lightcurve revealed a weak eclipse of the donating sdB star by a massive WD accretor, implying an accretor temperature of over $60,000 \, \rm K$ (see \citet{Kupfer2020b} for further details).

\section{Discussion} \label{sec:discussion}

\subsection{Selection Biases}

In order to determine what underlying population of objects our sample represents, we must consider selection biases introduced when identifying objects based on photometric periodicity and color. By selecting objects with the colors satisfying Pan-STARRS $(g-r)>0.2$ and $(r-i)>0.2$, our survey probes a large color space, effectively selecting all objects with $T>6,700\,\rm K$. Because our selection was a simple color cut, the objects in our sample span a large dynamic range in absolute luminosity, ranging from DWDs to systems containing helium stars. The selection does omit cool objects ($T<6,700\,\rm K$), and cool secondary of ZTF J1539+5027 suggests the possibility that tides may not efficiently heat all WDs at short orbital periods \citep{Burdge2019a}, suggesting that there may be value in extending the selection to encompass a cooler sample of objects as well. 

Binary systems containing a high temperature component are prone to exhibiting pronounced eclipses, and are intrinsically more luminous than their cool counterparts, allowing us to probe a larger volume on average by targeting blue systems than if we simply selected objects at all colors. Less luminous cool WDs do outnumber hot WDs, meaning that the space density of short period DWDs could be substantially larger than that of the high temperature systems we are targeting if WDs in close binaries are not efficiently heated by tides. We believe our omission of cool WDs is unlikely to have significantly diminished the number of DWDs detectable in ZTF, because WDs are intrinsically faint due to their small size, and cool WDs fall below the ZTF's detection threshold at much closer distances than their hot counterparts, greatly limiting the volume which can be probed in the cool temperature regime (consider a $0.3\,  M_\odot$ He WD at $6,700\rm \, K$--such an object would have an apparent magnitude of $21$ in $g$ at a distance of just $430 \rm \, pc$, and a $0.6\,  M_\odot$ CO core WD at this temperature would reach this apparent magnitude at just $270 \rm \, pc$). 

We also searched the entire catalog of WD candidates reported in \citet{Fusillo2019}, which includes cooler WDs, and discovered no binary candidates which were not already encompassed by the Pan-STARRS color cut. Of the 15 systems described here, ZTF J0538+1953, PTF J0533+0209, ZTF J0722$-$1839, ZTF J2228+4949, ZTF J0640+1738, ZTF J1901+5309, and ZTF J2320+3750 are members of the \citet{Fusillo2019} WD catalog. 

In contrast to optical surveys, gravitational wave surveys like \emph{LISA} are agnostic to the temperatures of sources, and thus will probe this cooler population of binaries, which remain hidden to optical surveys. Sources with luminous non-degenerate helium star components such as ZTF J0643+0318, ZTF J2130+4420, and ZTF J2055+4651 are intrinsically luminous enough that they can exceed ZTF's detection threshold at several kpc, even when heavily extincted, as is the case for both ZTF J0643+0318 and ZTF J2055+4651. Thus, by targeting redder sources, one could potentially expect to find more such systems.

We did not employ any selection invoking the astrometric solutions of the \emph{Gaia} survey \citep{Gaia2018} because ZTF's limiting apparent magnitude of $\approx 21$ is beyond the threshold of \emph{Gaia}'s ability to measure reliable parallaxes. Additionally, ZTF has acquired a significant amount of photometry in dense regions of the Galactic plane, where \emph{Gaia} astrometric solutions are less reliable than in lower density, higher galactic latitude fields. 

Curiously, of the eclipsing systems in our sample, only one contains a clear $>0.5\,M_{\odot}$ WD, even though the ELM survey found primarily pairs of He core+CO core WDs. However, this is easily understood as a selection effect. In general, CO WDs are less luminous than their He WD counterparts due to their smaller size. Thus, when such systems do undergo eclipses, it results in a shallow eclipse, as the CO WD subtends a small cross section of the larger He WD companion when it occults it, and during the secondary eclipse when the CO WD is occulted, little luminosity is lost, as the CO WD only contributes a small fraction of the total luminosity of the system (unless the CO WD is significantly hotter than the He WD, as is the case in ZTF J1539+5027).

Objects discovered via ellipsoidal modulation carry a different selection effect. As discussed in \citet{Faulkner1972}, there is a direct relation between the density of an object and the orbital period at which it overflows its Roche lobe, with only a weak dependence on mass ratio. In cases where the companion has a higher mass than the object overflowing its Roche lobe ($q<1$, as is necessarily the case for most double degenerate objects exhibiting ellipsoidal modulation), one can relate the orbital period at which the object fills its Roche lobe, $P_b$, in days, and the density $\rho$ of the Roche filling object in $g\,cm^{-3}$ with the useful approximation
\begin{equation}
    \label{eq:Period}
    P_b\approx \frac{0.43}{\rho^{1/2}}~,
\end{equation}
which does not deviate by more than three percent for $1<q<100$ \citep{Eggleton1983}. The selection effect introduced by this relation is quite apparent~--photometric surveys like ZTF only detect ellipsoidal modulation in objects nearly filling their Roche lobes, and therefore can only detect such binaries in a narrow range of orbital periods, governed by the lower density component in the system. Consequentially, high density objects such as CO WDs can only be discovered at extremely short orbital periods using ellipsoidal modulation, as these objects do not begin to fill their Roche lobes until $P_b \approx 1 \rm \, min$. At these orbital periods, gravitational radiation acts so quickly that the merger timescale of the systems is short (less than $500 \rm \, yr$ for a pair of $0.6\,M_{\odot}$ CO WDs at $60$\,s orbital period, which would exhibit ellipsoidal modulation on the order of a few percent at this period). However, selecting targets which exhibit ellipsoidal modulation does not introduce significant bias regarding the mass of the unseen companion, other than the basic requirement that it should be more dense/less luminous than the detected object (otherwise it would be the object dominating the signal). This has already manifested itself in our small sample of objects, as several of the systems exhibiting ellipsoidal modulation contain higher mass presumably CO core WDs, such as ZTF J2320+3750 and ZTF J0533+0209, whereas the majority of the eclipsing systems containing pairs of He WDs.

\begin{figure*}[htpb]
\begin{center}
\includegraphics[width=1.0\textwidth]{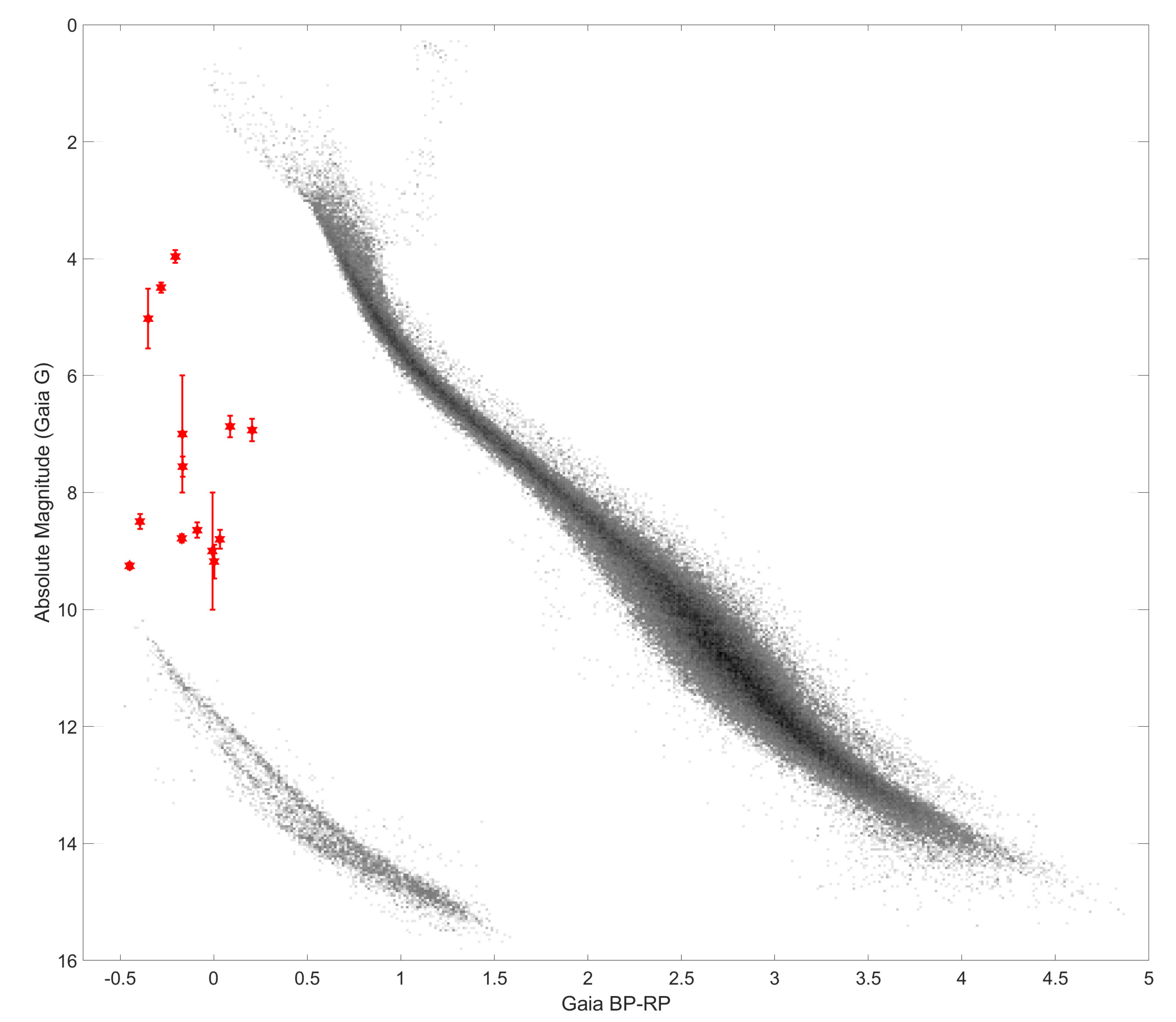}
\caption{Hertzsprung-Russell diagram illustrating the de-reddened locations of 14 binaries in the sample with \emph{Gaia} astrometric solutions (ZTF J0643+0318 is omitted, as its modelling is ongoing). The red stars represent objects which are in our sample, with absolute luminosities calculated based on their spectroscopic distances. Most objects cluster between absolute magnitudes of 6.5 and 10.0, with the exception of the systems containing either He burning stars or young and hot He WDs (which contribute significant additional luminosity, dwarfing both the luminosity of the companion WD and any accretion luminosity). The background color magnitude diagram (CMD) is the a sample of all stars in \emph{Gaia} within 100 parsecs that have reliable astrometric solutions.}
\label{fig:HR}
\end{center}
\end{figure*}  

In conclusion, it is unsurprising, based on selection biases, that our sample contains systems with at least one low-mass He WD, or a non-degenerate He star component, as detecting ellipsoidal modulation and/or eclipses is challenging for systems containing only CO WDs or more compact objects. Notably, the location of our sample when plotted on a Hertzsprung-Russell diagram (see Figure \ref{fig:HR}) reflects this, as all systems are clearly over-luminous relative to the WD cooling track. It is worth noting that detecting objects via photometry has important advantages over spectroscopic selection in that obtaining spectra requires dividing photons across many pixels, and thus requires integrating for much longer to exceed the signal per pixel necessary to avoid significant dilution of the SNR due to readout noise. This means that spectroscopic surveys must either target significantly brighter objects than a photometric survey, or sacrifice critical temporal resolution by acquiring longer exposures than a photometric survey of equivalent depth. Such considerations are important when comparing massively multiplexed spectroscopic surveys and photometric surveys like ZTF, and their discovery potential in the phase-space of short orbital period binaries.

\subsection{\emph{LISA}}

\begin{figure*}[htpb]
\begin{center}
\includegraphics[width=1.0\textwidth]{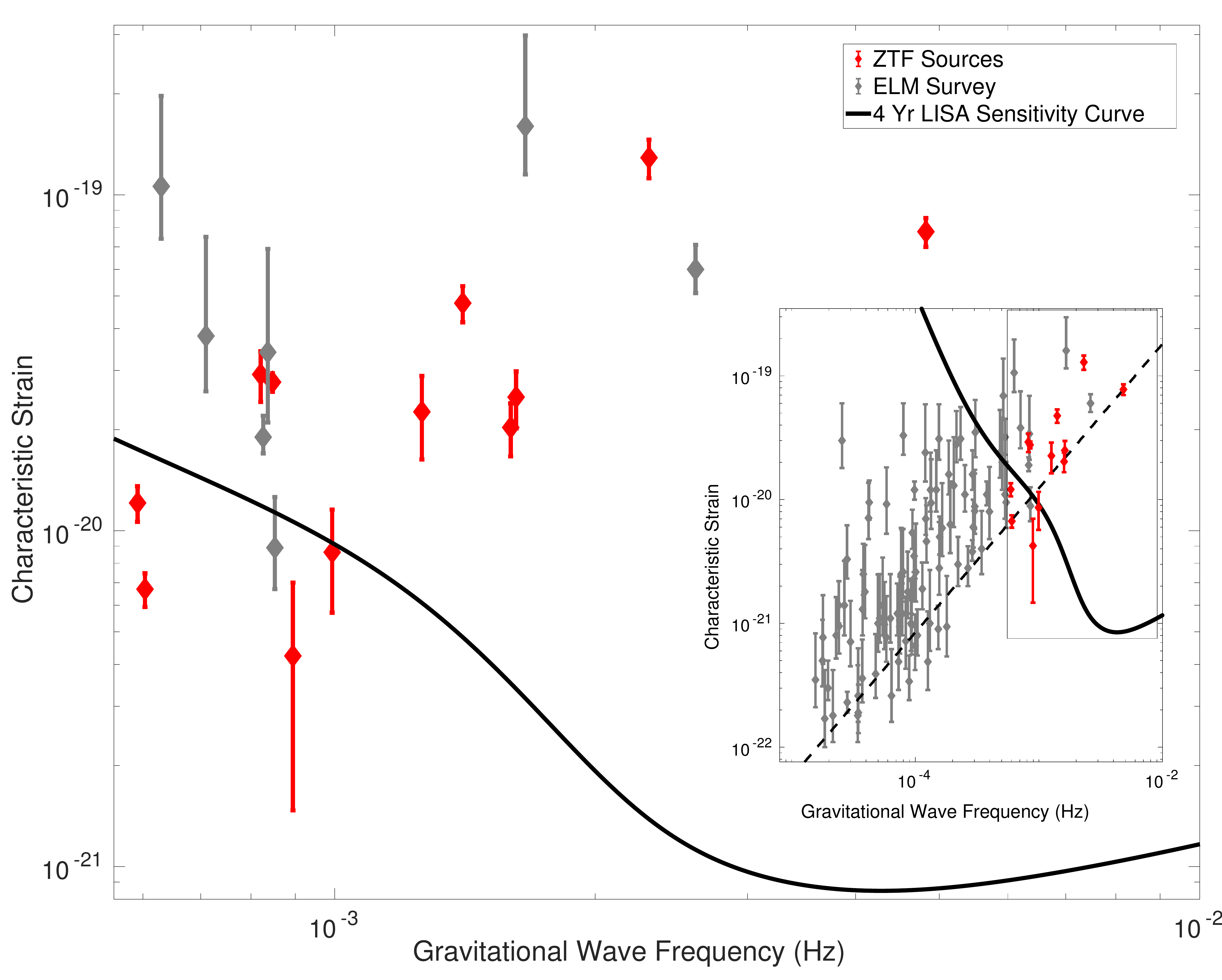}
\caption{The characteristic gravitational wave strain of the sources described in this work (shown as red diamonds--note that the three mass-transferring sources for which we do not yet have parameter estimates have been omitted), compared with the sample of DWD binaries from the ELM survey (shown as grey diamonds), overplotted with the 4 yr \emph{LISA} sensitivity curve (solid black line). An inset compares the sample presented in this paper (which exclusively investigated systems with orbital periods under an hour), with the full sample of systems discovered by the ELM survey \citep{ELMVIII}. The inset also presents the evolution of ZTF J1539+5027 with the dashed black line, illustrating what its characteristic strain would have been in the past at longer orbital periods. Overall, the ZTF sample contains two high SNR gravitational wave sources (ZTF J1539+5027 and ZTF J0538+1953, with 4-year SNRs of approximately 96). See Table \ref{tab:LISA} for further details. The 4 yr \emph{LISA} sensitivity curve was computed as described in \citet{Robson2019}.}
\label{fig:LISA}
\end{center}
\end{figure*}

Our survey targets binary systems with orbital periods under an hour because \emph{LISA}'s gravitational wave sensitivity peaks in this regime. As illustrated in Figure \ref{fig:LISA}, half of our sample falls above the \emph{LISA} sensitivity curve after 4 years of observations, though the current uncertainty in \emph{LISA} SNR for most sources originates from an uncertainty in the distance to the sources. It is also worth noting that the characteristic strains plotted in Figure \ref{fig:LISA} are a quantity which does not account for inclination, and as discussed earlier, a near edge-on inclination significantly diminishes the gravitational wave signal expected from a binary compared to a face-on system. Many of the ZTF sources consist of two He core WDs, whereas the ELM survey is dominated by He WDs orbiting higher mass CO core counterparts, with the notable exception of the system reported in \citet{Brown2020b}, meaning that on average the chirp mass is lower in the ZTF sample of objects than in the ELM survey. 

The average orbital period of the systems in the ZTF sample is significantly shorter than in the ELM sample, due both to selection criteria (this work did not pursue objects with $P_{b}>60 \rm \, min$), and selection bias, as eclipses and ellipsoidal modulation become more difficult to detect at longer periods, though future work will characterize longer period systems. Shorter period objects like ZTF J1539+5027 have a small lifetime (see Table \ref{tab:LISA}) and are thus rarer, and likely to be more distant if detected at all. 

As illustrated in Figure \ref{fig:LISA}, targeting objects at short periods compensates for the loss in characteristic strain due to such systems being rarer (and thus more distant), as LISA's sensitivity increases by more than an order of magnitude over the period range of objects in our sample. The increase in sensitivity from \emph{LISA} couples with the $f^{2/3}$ frequency dependence of the gravitational wave amplitude, and the boost in SNR due to $\sqrt{N_{cycle}}$ over 4 years of observations, meaning that the characteristic strain scales as $f^{7/6}$. Thus, a system like ZTF J1539+5027 has an enormous advantage in detectability by \emph{LISA} over a higher chirp mass system such as J2055+4651.

The most significant contributions of our survey to the sample of \emph{LISA}-detectable binary systems are ZTF J1539+5027 and ZTF 0538+1953, which along with SDSS J0651+2844 form a trio of high SNR eclipsing \emph{LISA}-detectable binaries. Due to their eclipses, we can precisely measure time these systems using optical data, and thus given a few months to years of monitoring, measure orbital decay with high confidence, and potentially even measure the acceleration of orbital decay. 

Many of our binaries have precisely constrained inclinations as a result of features in their lightcurves (in particular those systems exhibiting eclipses). Eventually, these systems will serve as a test of \emph{LISA}, which can also measure inclinations in systems through the ratio of the amplitudes of the two gravitational wave polarizations, which have differing inclination dependence. The cross polarization, $h_{\times}$, scales as $(1+\cos^2(i))$, whereas the plus polarization, $h_{+}$, scales as $\cos(i)$, thus vanishing for the edge-on inclination which most eclipsing systems are near. Ellipsoidal variables like PTF J0533+0209 stand to benefit significantly from a \emph{LISA} gravitational wave signal, which can be used to precisely constrain the inclination of these systems. When combined with the radial velocity semi-amplitude and photometric ellipsoidal modulation amplitude, this allows for a robust estimate of the component masses.

When \emph{LISA} begins to operate, we will gain new information on the sources. For a system like ZTF J1539+5027, which is already well constrained electromagnetically (but also a high SNR \emph{LISA} source), the primary new constraint will be a precise distance estimate. Currently, this estimate is based on the measured radius and temperature of the object (known as a spectroscopic distance estimate), but such an estimate is quite model dependent, and assumes a reliable understanding of variables such as extinction along the line of sight. \emph{LISA} will provide a precisely measured gravitational wave strain amplitude, and the chirp mass is already well estimated for this system based on electromagnetic constraints, and thus one can infer the distance to the source using this amplitude. For some sources, particularly those which are ellipsoidal variables with well measured temperatures, this precise distance measurement will help further constrain the properties of the system by placing tight constraints on the optical luminosity of the system.

Perhaps one of the most exciting class of systems \emph{LISA} will detect are those like ZTF J1539+5027, but also bright enough to have their distances measured precisely via parallax. With such sources, one could use the gravitational-wave strain amplitude to precisely measure the chirp mass of the system in a robust manner. Such a measurement would be extremely exciting, as these systems have precisely measurable orbital decay rates from eclipse timing (or even if they lack optical periodicity to measure this from, \emph{LISA} could be used to measure the frequency evolution of the system). One could use such systems to directly measure the difference in measured $\dot{P_b}$ from that expected due to general relativity, probing the efficiency of tides in the system. Currently, the only system in our sample which has a distance measured precisely enough for such an exercise is ZTF J2130+4420, which unfortunately is a rather low SNR \emph{LISA} source. Additionally, this system is undergoing mass transfer, so any measured deviation of the orbital evolution from general relativity could be influenced by an exchange of angular momentum due to mass transfer.

\begin{deluxetable*}{ccccccc}[htbp]
\tablenum{6}
\tablecaption{LISA SNR of systems \label{tab:LISA}}
\tablewidth{0pt}
\tablehead{Name & \colhead{4 yr \emph{LISA} SNR} & \colhead{$\mathcal{A}$}& \colhead{$2\cos(i)$}& \colhead{$1+\cos^2(i)$}  & \colhead{Decay Timescale (Myr)}}

\startdata
ZTF J1539+5027 & $\sim 96$& $(9.96\pm0.93)\times 10^{-23}$ & $0.204\pm0.022$ & $1.0104\pm0.0023$ & $0.20753\pm0.00043$  \\
ZTF J0538+1953 &  $\sim 96$ & $(2.38\pm0.30)\times 10^{-22}$ & $0.1593\pm0.0031$ & $1.00635\pm0.00024$ & $1.42\pm0.17$ \\
PTF J0533+0209 &  $\sim 8$  & $(5.5\pm1.1)\times 10^{-23}$ & $0.591\pm0.046$ & $1.087\pm0.014$ &$3.89\pm0.95$ \\
ZTF J2029+1534 & $\sim 5$  & $(4.52\pm0.78)\times 10^{-23}$ & $0.117\pm0.024$ & $1.0034\pm0.0014$ & $5.35\pm0.85$ \\
ZTF J0722$-$1839 &$\sim 8$  & $(1.13\pm0.14)\times 10^{-22}$ & $0.0119\pm0.0076$ & $1.000036\pm0.000050$ &  $5.95\pm0.71$ \\
ZTF J1749+0924 &$\sim 3$  & $(5.7\pm1.5)\times 10^{-23}$ & $0.158\pm0.049$ & $1.0062\pm0.0039$ &  $9.1\pm2.1$ \\
ZTF J1946+3203 &$<1$ & $(2.46\pm0.80)\times 10^{-23}$ & $0.447\pm0.055$ & $1.050\pm0.012$ &  $23\pm11$ \\
ZTF J0640+1738 &$<1$& $(9.2\pm3.8)\times 10^{-24}$ & $0.83\pm0.19$ & $1.174\pm0.079$  & \nodata \\
ZTF J2130+4420 &$\sim 2$& $(8.47\pm0.53)\times 10^{-23}$ & $0.127\pm0.035$ & $1.0040\pm0.0022$ & $16.6\pm8.1$ \\
ZTF J1901+5309 & $\sim 2$  & $(9.05\pm0.15)\times 10^{-23}$ & $0.095\pm0.017$ & $1.00226\pm0.00083$ & $24.5\pm3.8$  \\
ZTF J2320+3750 &$<1$ & $(1.58\pm0.18)\times 10^{-23}$ & $0.19\pm0.11$ & $1.009\pm0.012$ &  $55.8\pm3.1$  \\
ZTF J2055+4651 & $<1$ & $(4.42\pm0.51)\times 10^{-23}$ & $0.233\pm0.017$ & $1.0136\pm0.0020$ & $30.9\pm3.3$  \\
\enddata
\tablecomments{The sky and inclination dependent estimated \emph{LISA} SNR of all systems in our sample (we marginalize over polarization angle), after 4 years of observations. All values are calculated using spectroscopic distances (see Table \ref{tab:distances}), and the characteristic decay timescale assuming gravitational wave decay, defined as $\frac{3}{8}\frac{P_{b}}{|{\dot P_{b}|}}$. As a point of comparison, we estimate SDSS J0651+2844 will reach an SNR of $\sim 88$ after four years (with $M_{A}=0.49\,  M_\odot\ $, $M_{B}=0.247\,  M_\odot\ $, and $d=0.933\,\rm  kpc $). We have omitted estimates for the mass-transferring systems for which we do not have parameter estimates, and have ommitted a decay timescale estimate for ZTF J0640+1738 due to uncertainties larger than the estimate. The gravitational wave amplitude is defined as $\mathcal{A} = \frac{2(G\mathcal{M})^{5/3}}{c^4d}(\pi f)^{2/3}$, and the $2\cos(i)$ and $1+\cos^2(i)$ give the inclination dependent coefficients of the two gravitational wave polarizations, $h_{+}$ and $h_{\times}$, respectively. }
\end{deluxetable*}

\subsection{Evolutionary Fates}

\begin{figure*}[htpb]
\begin{center}
\includegraphics[width=1.0\textwidth]{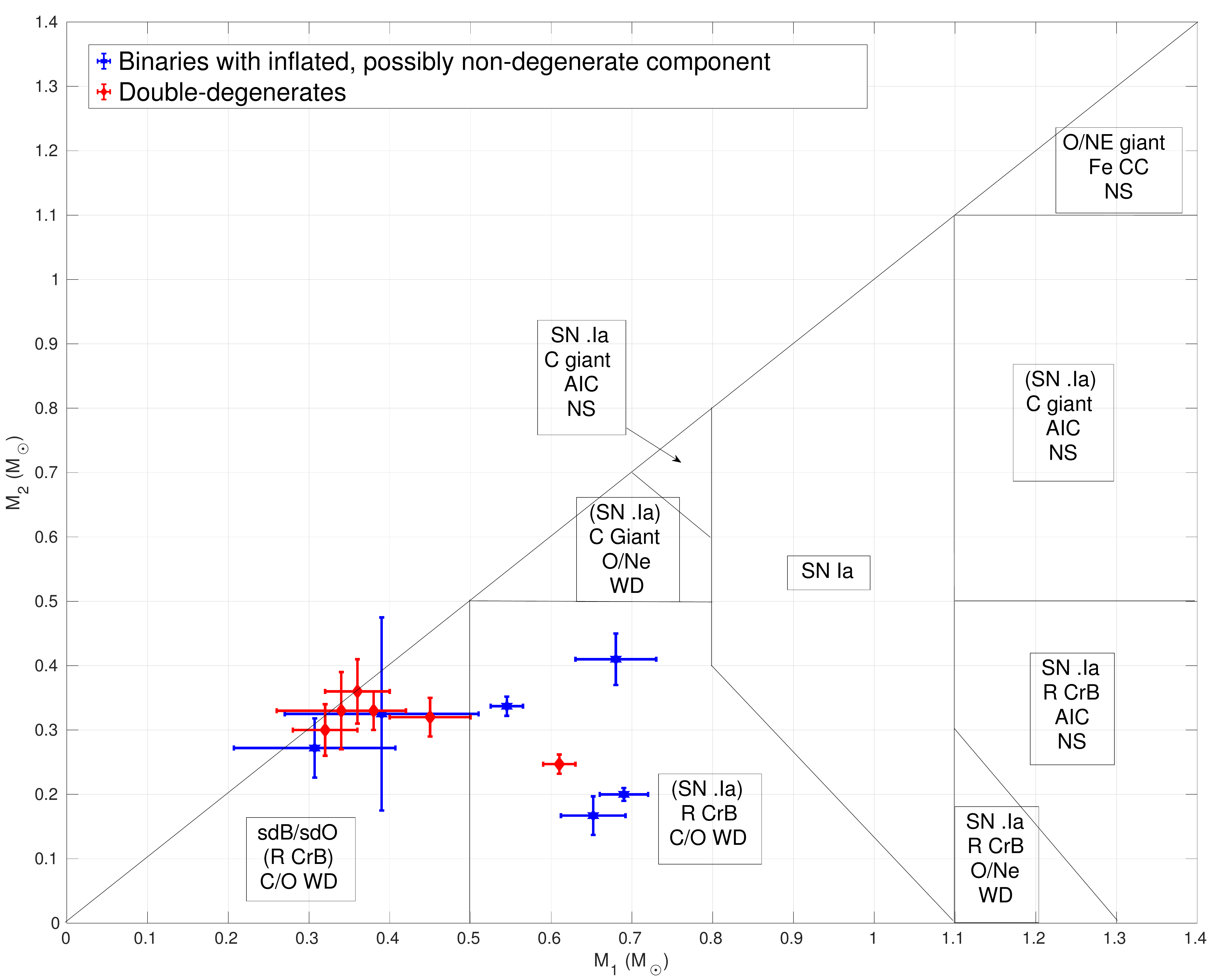}
\caption{A reproduction of the diagram featured in \citet{Shen2015}, illustrating possible evolutionary fates of double degenerate pairs. In red, we plot the double degenerate objects discovered by our survey. Most of these are double He WDs that are expected to evolve into hot subdwarf stars (sdB/sdO) objects upon merger. Three other double degenerate systems (including ZTF J1539+5027) have CO cores and likely produce R CrB stars upon merger, though they could also turn into mass transferring AM CVn systems that exhibit SN .Ia He-shell detonations. Systems containing non-degenerate objects (blue points) would only follow the scenarios described in \citet{Shen2015} if the He-burning objects transitioned to a degenerate state prior to merger. They could produce other phenomena, including type Ia SNe, due to mass transfer/merger before becoming degenerate (see text).}
\label{fig:Shen}
\end{center}
\end{figure*}

All of the binaries in our sample are either already interacting, or will interact in the near astronomical future (see Table \ref{tab:LISA}). Thus, a significant question worth posing is what will occur when these systems interact. We illustrate possible outcomes should the systems merge in Figure \ref{fig:Shen}, based on the work described in \citet{Shen2015}.

Most of the double He WD systems in our sample are close to a mass ratio of unity, and thus are likely to merge when they interact \citep{Marsh2004}. As discussed in \citet{Shen2015}, these systems could evolve into two classes of objects: either a non-degenerate He-burning hot subdwarf star, which eventually cools into a CO WD, or alternatively, they could form an R CrB star, which would also eventually cool into a CO WD. In either case, these systems ultimately end up on the WD track, and thus serve as evidence that it is likely at least some low mass WDs merge and ultimately form a CO WD.

The other double degenerate systems in the sample, containing both a CO WD and a He WD, have larger mass ratios than double He WD systems. Thus, with the onset of mass transfer, it is possible that these systems could evolve into AM CVns \citep{Ramsay2018}. However, \citet{Shen2015} proposes that these systems too could ultimately merge, and along the way potentially produce powerful helium detonations known as .Ia (``dot Ia'') supernovae. Upon merger, they are expected to form an R CrB as the He WD is disrupted by its more dense CO companion, forming a helium atmosphere around this core. Ultimately, most of these R CrB stars would cool to form a CO WD on the WD cooling track.

Finally, the remaining systems in our sample contain a degenerate WD accreting from a non-degenerate He-burning star (ZTF J2130+4420 and ZTF J2055+4651). These systems will likely build up a layer of helium on the CO WD, which could eventually undergo a detonation, and if the CO WD is sufficiently massive, this could result in a ``double-detonation'' by igniting the degenerate carbon-oxygen core \citep{Shen2018a,Shen2018b}. However, in the case of ZTF J2130+4420, the CO WD is only $0.545^{+0.020}_{-0.020}\,M_{\odot}$, likely too low even to produce a sub-luminous Type Ia supernova, whereas ZTF J2055+4651, with its $0.68^{+0.05}_{-0.05}\,M_{\odot}$, is the more likely of the two objects to produce such an event \citep{Perets2019}. Ultimately, the two objects could also exhaust helium shell burning, and simply cool into double degenerates. Upon merger, they would likely form a rapidly rotating CO WD, potentially preceded by an R CrB phase if substantial He remains as expected for hybrid WDs \citep{Perets2019}. For further details on these systems, please see \citet{Kupfer2020a,Kupfer2020b}.

\section{Conclusion} \label{sec:conclusion}

Using data from the Zwicky Transient Facility, we have significantly increased the number of known binary systems with an orbital period of less than an hour. Using high-speed photometric follow-up in combination with spectroscopic follow-up we have characterized the physical parameters of the systems using model-dependent techniques invoking WD mass-radius relations. In future work, we hope to obtain more model-independent measurements such as orbital decay rates to more robustly characterize the physical parameters of these systems.

We will continue to analyze ZTF data as it accumulates more epochs and discover more such objects. At the current stage of the survey, the sample we have discovered exhibits a remarkably high fraction of pairs of He WDs, likely due to selection effects. As we continue the survey, we anticipate discovering more systems, and that our discoveries will begin to include other classes of sources, such as pairs of CO WDs. All of the detached systems in our sample should be undergoing rapid orbital decay due to general relativity, and we expect to detect this orbital decay in every detached system with high significance before \emph{LISA} launches. 

The algorithms and analysis techniques we are developing will be applicable to upcoming surveys such as the one which will be conducted by the Vera Rubin Observatory \citep{LSST2009}. These data sets will be accompanied by their own challenges (in particular, lower cadence, split across many filters). Given the clear promise demonstrated by the discoveries emerging from ZTF, we are optimistic that optical time domain surveys will continue to revolutionize our understanding of binaries in the millihertz regime. These surveys will set the stage for \emph{LISA}, having discovered dozens, if not hundreds of sources by the time the instrument begins to operate. In future work, we will characterize our recovery efficiency in greater detail using synthetic lightcurves, in order to better characterize ZTF's sensitivity to detecting the population of DWDs in the galaxy. Efforts are currently underway to increase sensitivity to the shortest period binary systems by account for their orbital period evolution, which requires algorithms that can accommodate acceleration searches \citep{Katz2020}.

Binaries with compact object components are transitioning into a golden age, as we enter an era when time domain surveys such as ZTF enable detection via periodic features. Massively multiplexed spectroscopic surveys acquiring multiple epochs such as SDSS-V \citep{Kollmeier2017} and LAMOST \citep{Cui2012} will enable detection of these systems via large radial velocity shifts, with the feasibility of this technique having already been demonstrated on a smaller scale by the ELM survey. Deep all sky X-ray surveys such as that being conducted by the eROSITA instrument aboard the SRG mission \citep{Merloni2012} will enable detection of ultracompact X-ray binaries and direct impact accreting systems such as HM Cancri \citep{Roelofs2010}. Finally, \emph{Gaia} \citep{Gaia2018} has completely revolutionized stellar astronomy by providing parallax measurements, which has enabled careful targeting of objects such as WDs. This era will culminate with the launch of \emph{LISA}, which will enable detection of thousands of these systems in the form of gravitational waves, and revolutionize our understanding of compact binary systems. 

\acknowledgments
K.B.B thanks the National Aeronautics and Space Administration and the Heising Simons Foundation for supporting his research.

M.~W.~C. acknowledges support from the National Science Foundation with grant number PHY-2010970.

P.R-G acknowledges support from the State Research Agency (AEI) of the Spanish Ministry of Science, Innovation and Universities (MCIU), and the European Regional Development Fund (FEDER) under grant AYA2017--83383--P.

V.S.D, ULTRACAM and HiPERCAM are supported by STFC.

The research leading to these results has received funding from the European Research Council under the European Union's Horizon 2020 research and innovation programme numbers 677706 (WD3D) and 340040 (HiPERCAM).

Based on observations obtained with the Samuel Oschin Telescope 48-inch and the 60-inch Telescope at the Palomar Observatory as part of the Zwicky Transient Facility project. ZTF is supported by the National Science Foundation under Grant No. AST-1440341 and a collaboration including Caltech, IPAC, the Weizmann Institute for Science, the Oskar Klein Center at Stockholm University, the University of Maryland, the University of Washington, Deutsches Elektronen-Synchrotron and Humboldt University, Los Alamos National Laboratories, the TANGO Consortium of Taiwan, the University of Wisconsin at Milwaukee, and Lawrence Berkeley National Laboratories. Operations are conducted by COO, IPAC, and UW.

Based on observations made with the Gran Telescopio Canarias (GTC) installed in the Spanish Observatorio del Roque de los Muchachos of the Instituto de Astrof\'isica de Canarias, in the island of La Palma.

Based on observations made at the European Southern Observatory New Technology Telescope (NTT), La Silla.

The KPED team thanks the National Science Foundation and the National Optical Astronomical Observatory for making the Kitt Peak 2.1-m telescope available. The KPED team thanks the National Science Foundation, the National Optical Astronomical Observatory and the Murty family for support in the building and operation of KPED. In addition, they thank the CHIMERA project for use of the Electron Multiplying CCD (EMCCD).

Some of the data presented herein were obtained at the W.M. Keck Observatory, which is operated as a scientific partnership among the California Institute of Technology, the University of California and the National Aeronautics and Space Administration. The Observatory was made possible by the generous financial support of the W.M. Keck Foundation. The authors wish to recognize and acknowledge the very significant cultural role and reverence that the summit of Mauna Kea has always had within the indigenous Hawaiian community. We are most fortunate to have the opportunity to conduct observations from this mountain.

This article is based on observations made in the Observatorios de Canarias del IAC with the William Herschel Telescope operated on the island of La Palma by the Isaac Newtoun Group of Telescopes in the Observatorio del Roque de los Muchachos.

This research benefited from interactions at the ZTF Theory Network Meeting that were funded by the Gordon and Betty Moore Foundation through Grant GBMF5076 and support from the National Science Foundation through PHY-1748958.

\vspace{5mm}
\facilities{PO:1.2m (ZTF), KPNO:2.1m (KPED), NTT (ULTRACAM), Hale (CHIMERA, DBSP), GTC (HiPERCAM), Keck:I (LRIS)}


\bibliography{sample63}{}
\bibliographystyle{aasjournal}

\appendix

\section{Instrumental/Observational Information}

\subsection{LRIS}

We obtained the majority of our spectroscopic follow-up using the Low Resolution Imaging Spectrometer (LRIS) \citep{Oke1995} on the 10-m W.\ M.\ Keck I Telescope on Mauna Kea. We conducted most of our observations using the 600 grooves/mm grism with a blazing angle of 4000 $\rm \AA$, and the 1 arcsecond slit on the spectrograph, resulting in an effective FWHM resolution of 3.8-4.1 across the wavelength range of the blue arm (the approximate wavelength range is 3200--5800 $\rm \AA$). The dispersion of this element results in approximately 0.63 $\rm \AA$ of wavelength per pixel, meaning that we sample the PSF from this slit with more than 6 unbinned pixels, thus, we always bin along the dispersion axis when observing with this instrument, as it significantly reduces readout time and read noise, and comes with no cost in resolving power (at least in the case of 2x2 binning). For some observations which required particularly short exposures (such as ZTF J1539+5027), we binned 4x4, to further reduce readout time. Ultimately, the lower limit of exposure times during spectroscopic observations are set by the readout duty cycle and read noise floor (one splits photons across many pixels when obtaining spectra, resulting in a small number of counts per pixel, which can quickly become comparable to the readout noise in short exposures). Frame transfer CCDs offer an attractive solution to eliminating the readout duty cycle bottleneck (which, for observing ZTF J1539+5027 for example, was a prohibitive 34 percent, as the effective exposure time was 52 seconds, but even with 4x4 binning, the readout time was still 27 seconds). Electron Multiplying CCDs take this one step further by offering both the possibility of eliminating the issue of readout time, and also eliminating read noise, meaning that observations can be obtained in photon-counting mode. The catch of such an instrument is a factor of $\sqrt{2}$ larger readout noise due to concidence losses \citep{Tulloch2011}; however, for faint targets which require high temporal resolution, these CCDs offer a solution that conventional CCDs do not. Coincidence losses can be avoided in the photon-counting limit where no more than one photon is incident on a pixel in any given image, making this technology particularly attractive for higher resolution CCDs where the dispersion is sufficiently high that this regime can be easily reached.

\subsection{DBSP}

We obtained the spectrum of ZTF J2228+4949 using the Double-Beam Spectrograph (DBSP) \citep{Oke1982} on the $200\,\rm inch$ Hale telescope at Palomar observatory. The instrument provides a resolution comparable to LRIS, and has similar limitations in readout time overheads.

\subsection{ISIS}
Spectra for ZTF J2130+4420 were obtained using the ISIS spectrograph on the $4.2\,\rm m$ William Herschel Telescope. For further details on these observations, please see \citet{Kupfer2020a}.

\subsection{GMOS-N}
Spectra for ZTF J2055+4651 were obtained using the GMOS-N spectrograph on the $8.1\,\rm m$ Gemini North Telescope. For further details on these observations, please see \citet{Kupfer2020b}.

\subsection{HiPERCAM}

HiPERCAM is a high speed photometer mounted on the 10.4-m Gran Telescopio Canarias (GTC) on the island of La Palma in the Canary Islands (\citealt{Dhillon2018}, Dhillon et al. in prep). The instrument is a frame transfer quintuple-beam imager, allowing for simultaneous imaging in $u$, $g$, $r$, $i$, and $z$ band at frame rates of $>1 \rm \, kHz$. The light collecting power of a 10 meter class telescope, combined with the high frame rate of this instrument, makes it optimal in obtaining the high SNR observations of the systems described in this work, particularly systems such as ZTF J1539+5027, whose eclipse ingress and egress last only a few seconds and spans an apparent magnitude range of approximately 21 in $g$ at the start of the eclipse, and reaches $>27$ in $g$ at the base of the eclipse. 

\subsection{ULTRACAM}

ULTRACAM is a three-channel high speed photometer currently mounted on the 3.5 meter New Technology Telescope at La Silla observatory \citep{Dhillon2007}. All observations on this instrument were conducted using $u$ and $g$ filters on two of the channels, and the remaining channel either $r$ or $i$ band. Like the other high speed photometers used in our observations, the instrument is operated in frame transfer mode, effectively eliminating the readout duty cycle.

\subsection{CHIMERA}

The Caltech HIgh-speed Multi-color camERA (CHIMERA) \citep{Harding2016} is a dual-channel high speed photometer mounted on the prime focus of the 200-inch Hale telescope at Palomar Observatory. CHIMERA consists of a pair of electron multiplying CCDs capable of using either conventional amplifiers or those with electron multiplying gain. We operated the detectors in the conventional amplifier mode with frame transfer enabled, as for all of our targets we obtained enough photons in several second exposures such that the $\sqrt{2}$ increase in shot noise from using the electron multiplying gain would erase any gains from eliminating readout noise.

\subsection{KPED}

The Kitt Peak Electron Multiplying CCD Demonstrator (KPED) is a single-channel high speed photometer mounted on the 2.1 meter telescope at Kitt Peak National observatory \citep{Coughlin2019}. The instrument uses the same electron multiplying CCDs used by CHIMERA. The observations highlighted in this work were obtained using frame transfer mode and in $g$.

\begin{deluxetable*}{ccccc}[htbp]
\tablenum{7}
\tablecaption{Observations of systems used in analyses in this publication \label{tab:Observations}}
\tablewidth{0pt}
\tablehead{Name & \colhead{Instrument}  & \colhead{Observation Dates (UTC)} & \colhead{Configuration}& \colhead{Frame Exptime}}

\startdata
ZTF J1539+5027 & LRIS & June 16, July 12, 13 2018 & 600 grooves/mm grism, 4x4 binning & $52 \rm \, s$  \\
 & HiPERCAM  & June 2, 4, 5 2019 & u, g, r, i, z & $6,\, 3,\, 3,\, 6,\, 9 \rm \, s$  \\
ZTF J0538+1953 & LRIS & Sept 3, 27 2019 & 600 grooves/mm grism, 4x4 binning & $90 \rm \, s$  \\
 & HiPERCAM  & Sept 4, 8 2019 & u, g, r, i, z & $3,\, 1,\, 1,\, 3,\, 4 \rm \, s$  \\
ZTF J1905+3134 & LRIS & July 5 2019 & 400 grooves/mm grism, 2x2 binning & $300 \rm \, s$  \\
 & KPED  & July 3 2019 & g & $10 \rm \, s$  \\
PTF J0533+0209 & LRIS & Nov 15 2017, Mar 19 2018 & 400 grooves/mm grism, 2x2 binning & $120 \rm \, s$  \\
 & CHIMERA  & Dec 14, 15 2017 & g, i & $10 \rm \, s$  \\
PTF J2029+1534 & LRIS & Mar 21 2020 & 600 grooves/mm grism, 2x2 binning & $120 \rm \, s$  \\
 & CHIMERA  & Oct 1 2019 & g, r & $3,\,3 \rm \, s$  \\
ZTF J0722$-$1839 & LRIS & April 5, 6 2019, Feb 17 2020 & 600 grooves/mm grism, 2x2 binning & $141 \rm \, s$  \\
 & ULTRACAM  & Jan 26 2020 & u, g, i & $12,\,6,\,6 \rm \, s$  \\
ZTF J1749+0924 & LRIS & Mar 21 2020 & 600 grooves/mm grism, 2x2 binning & $120 \rm \, s$  \\
 & CHIMERA & July 15 2020 & g, r & $5,\,5 \rm \, s$  \\
ZTF J2228+4949 & DBSP & July 31, 2019 & 600 grooves/mm grating  & $200 \rm \, s$  \\
ZTF J1946+3203 & LRIS & Sept 27 2019 & 600 grooves/mm grism, 2x2 binning & $168 \rm \, s$  \\
 & CHIMERA  & August 6, 2019 & g, i & $3,\,3 \rm \, s$  \\
ZTF J0643+0318 & LRIS & Jan 25, 2020 & 600 grooves/mm grism, 2x2 binning & $170 \rm \, s$  \\
 & ULTRACAM  & September 28, 2019 & u, g, i & $24,\,8,\,8 \rm \, s$  \\
ZTF J0640+1738 & LRIS & Feb 17 2019 & 600 grooves/mm grism, 2x2 binning & $157 \rm \, s$  \\
ZTF J2130+4420 & ISIS & June 25, 26 2019 & R300B grating  & $120 \rm \, s$  \\
 & HiPERCAM  & July 8 2019 & u, g, r, i, z & $3.54,\, 1.77,\, 1.77,\, 1.77,\, 3.54 \rm \, s$  \\
ZTF J1901+5309 & LRIS & Sept 3 2019 & 600 grooves/mm grism, 2x2 binning & $141 \rm \, s$  \\
 & CHIMERA  & August 8 2019 & g, i & $3,\,3 \rm \, s$  \\
ZTF J2320+3750 & LRIS & Sept 27 2019 & 600 grooves/mm grism, 2x2 binning & $138 \rm \, s$  \\
ZTF J2055+4651 & GMOS-N & Sept 24, 25 2019 & B600 grating & $180 \rm \, s$  \\
 & HiPERCAM  & July 8 2019 & u, g, r, i, z & $9,\, 3,\, 3,\, 3,\, 9 \rm \, s$  \\
\enddata

\end{deluxetable*}

\begin{widetext}
\section{Period Finding}

In this section, we explore the sensitivity of the algorithms used to discover the systems described in this work using both a real example, as well as synthetic signals.

We typically search approximately 2 million trial frequencies per lightcurve. The highest frequency we search to is 720 times per day, and the lowest is defined as 2/baseline, where the baseline is the end date of the lightcurve, minus the start. Because the frequency grid depends on the baseline of the lightcurve, we compute it independently for each lightcurve. We used an oversampling factor of 4. To avoid aliases due to the sidereal day, we slice frequencies out of the frequency grid at 1.0 days, and its various harmonics.

One of the known systems we tested our technique on was the 12.75-min binary system, SDSS J0651+2844 \citep{Brown2011}, whose phase-folded ZTF lightcurve is illustrated in Figure \ref{fig:J0651}. This system did not become recoverable in the main ZTF survey until a substantial number of epochs had been accumulated (see Figure \ref{fig:period} for further details), as its eclipses are quite shallow due to the more luminous component being the larger. We expect that as ZTF's sampling continues to increase, we should start to recover more eclipsing CO WD plus He WD white dwarf systems like SDSS J0651+2844, whereas our current discoveries in general exhibit larger photometric amplitudes than seen in SDSS J0651+2844 because they consist primarily of pairs of He WDs, or as in the case of ZTF J1539+5027, the more compact CO WD dominates the luminosity.

In Figure \ref{fig:period} we illustrate the sensitivity of the conditional entropy algorithm to both a sinusoidal signal and a square wave as a function of both the amplitude of the signal and the number of epochs. In addition to characterizing the sensitivity of the algorithm to a synthetic signal, we also present ZTF's median RMS scatter as a function of apparent magnitude, and the cumulative ZTF sampling of \emph{Gaia} objects above a declination of $-28^{\circ}$ in Figure \ref{fig:period}. Using this figure, one can for example, infer that at an apparent magnitude of 19, ZTF has an RMS scatter of $\approx 7 \%$, and approximately two thirds of all sources in \emph{Gaia} above a declination of $-28^{\circ}$ have 500 or more detections in ZTF at this apparent magnitude. Using the top panels of the figure, one can then estimate that ZTF should have detected more than half of all sinusoidal sources at this apparent magnitude with an amplitude of $>3.5 \%$. These are crude estimates, and further work is underway to conduct careful simulations of ZTF data to better characterize detection efficiency by accounting for the cadence in each field, data artifacts, and other import elements that the simple treatment presented below cannot capture. The work presented in this publication is not intended to explore rates, and thus we leave the analysis at this.

\begin{figure}[htpb]
\begin{center}
\includegraphics[width=0.6\textwidth]{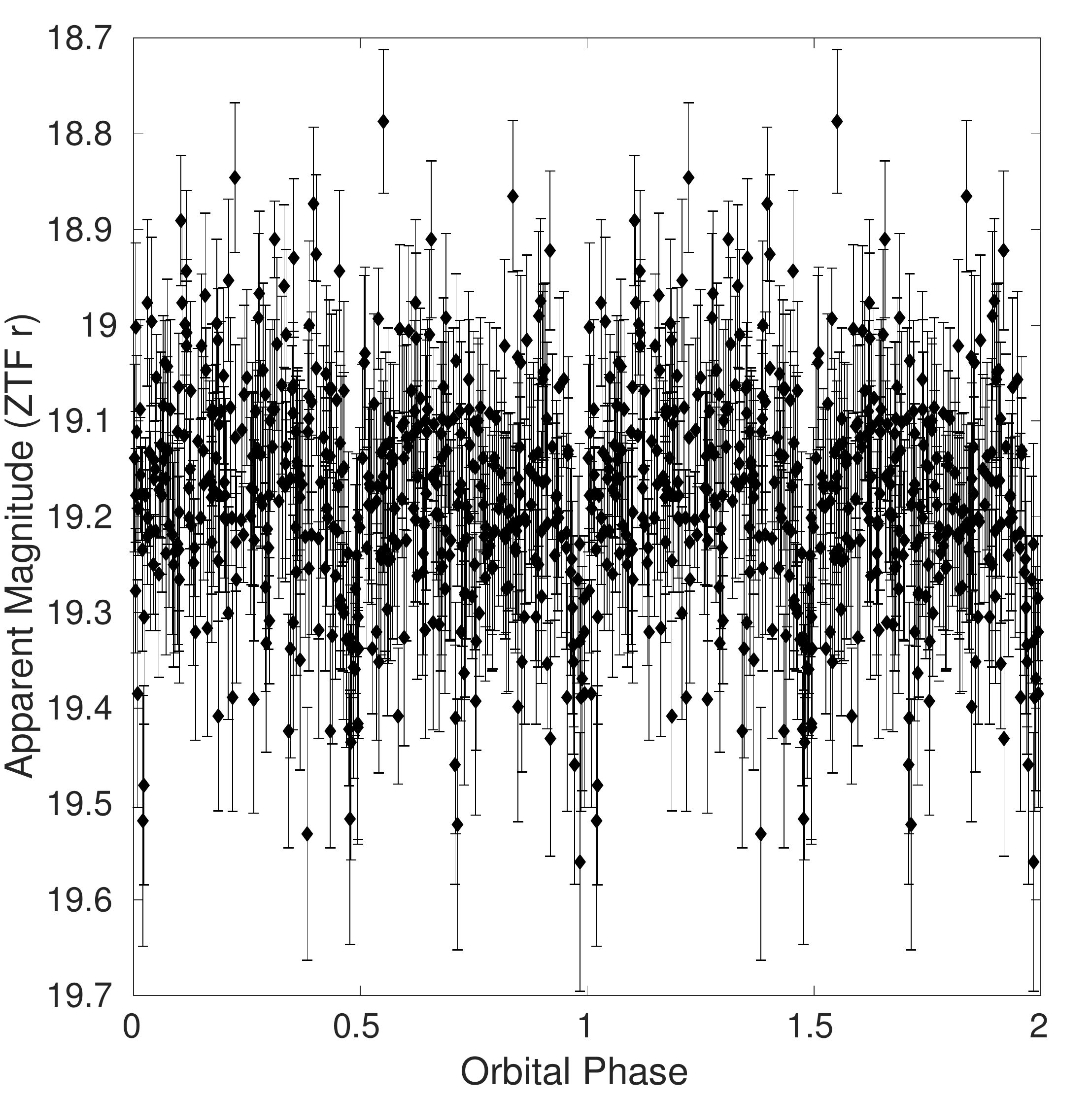}
\caption{The phase-folded ZTF lightcurve of the 12.75-min orbital period binary, SDSS J0651+2844 \citep{Brown2011}, which we blindly recover with our period finding algorithms. An early dedicated observation of this system during ZTF's commissioning served as a proof of concept that a wide field optical time domain survey such as ZTF could function as a powerful tool for identifying members of this class of gravitational wave source.}
\label{fig:J0651}
\end{center}
\end{figure}

\begin{figure*}[ht]
\begin{center}
\includegraphics[width=1.0\textwidth]{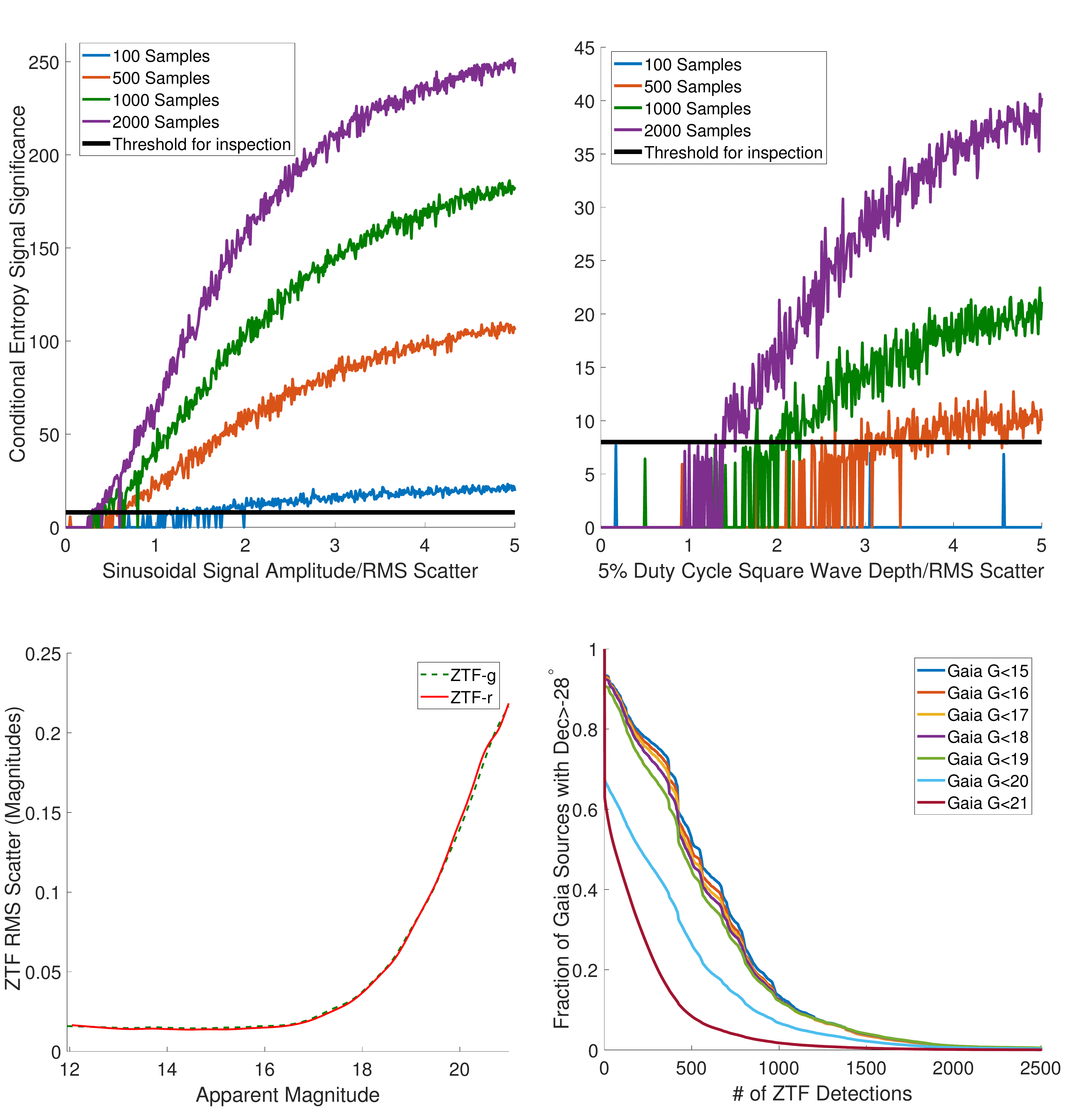}
\caption{Upper left: The significance returned by the conditional entropy algorithm on a synthetic sinusoidal signal with a 10 minute period. On the x axis, we illustrate the semi-amplitude of the sinusoid compared to the RMS scatter injected into the signal. The horizontal black line illustrates the significance threshold we imposed for inspecting lightcurves visually for candidates. Upper Right: This panel is analogous to the upper left panel, but rather than a synthetic sinusoidal signal, the algorithm was instead run on a synthetic eclipsing signal in the form of a periodic square wave with a 10 minute period, and 5 percent duty cycle. Lower left: An empirical average of ZTF's RMS scatter for a sample of all \emph{Gaia} sources with a color of $\rm (BP-RP)<0.5$. Lower right: An illustration of the fraction of \emph{Gaia} sources with a declination of $>-28^{\circ}$ and with a color of $\rm (BP-RP)<0.5$ vs the number of archival ZTF photometric detections of these sources. The large number of faint sources with no detections are dominated by sources which were not detected in the ZTF reference image used to seed the archival database (primarily located in dense regions of the Galactic plane). See \citet{Masci2019} for further details.}
\label{fig:period}
\end{center}
\end{figure*}  

\section{\emph{LISA} Signal to noise}

Here, we walk through the basic formalism we used to estimate the signal-to-noise ratios of the binaries in this paper.

In order to estimate the \emph{LISA} gravitational wave SNR, we adopt the same formalism as outlined in \citet{Robson2019}. In \citet{Burdge2019a} and \citet{Burdge2019b}, the chirp mass, $M_{c}$, was estimated from the rate of orbital decay due to gravitational wave emission. In this work, we do not yet have precise orbital decay measurements of any other systems, so we instead estimate the chirp mass from mass estimates based on lightcurve and spectroscopic modelling. We use the approximation for the orbit-averaged signal amplitude at the detector computed in \citet{Cornish2003},
\begin{equation}\label{eq:GWAmp}
   A^{2}=\mathcal{A}^2((1+\cos^2(i))^2\left\langle F^2_{+} \right\rangle+4 \cos^2(i)\left\langle F^2_{\times} \right\rangle),
\end{equation}where the intrinsic source amplitude is given by $\mathcal{A} = \frac{2(G\mathcal{M})^{5/3}}{c^4d}(\pi f)^{2/3}$, and the orbit averaged detector responses are $\left\langle F^2_{+} \right\rangle$ and $\left\langle F^2_{\times} \right\rangle$ (see \citet{Cornish2003} for the full expressions of these quantities). Note that \citet{Cornish2003} includes a factor of $\frac{1}{2}$ in this expression which we have omitted. This factor of $\frac{1}{2}$ arises from time averaging over the $|\cos{(\psi{t})}|^2$ in the signal; however, in order to find agreement with the sky, inclination, and polarization averaged SNR in \citet{Robson2019}, we found we had to eliminate this $\frac{1}{2}$. We believe this is because in \citet{Robson2019}, the phase of the gravitational wave is expressed as $e^{i\psi{f}}$, and when \citet{Robson2019} computes the inner product of the gravitational wave with itself, this term vanishes.

To estimate the signal-to-noise, we use the expression given in \citet{Korol2017},

\begin{equation}\label{eq:LISA_AMP}
   SNR^2=\frac{A^{2}T_{obs}}{P_{n}(f_{s})},
\end{equation}where $P_{n}(f_{s})$ is the power spectral density of the noise in a Michelson channel. We obtain the sky averaged \emph{LISA} noise curve from \citet{Robson2019} In order to estimate the power spectral density of the noise, we divide the sky averaged sensitivity by the sky averaged response function, $\sqrt{3/20}$ to obtain an effective non-sky averaged curve (see \citet{Robson2019} for further details). Note that while the response function given in \citet{Robson2019} is $\sqrt{3/10}$, it is actually a composite of a response which arises from averaging over the sky (the $\sqrt{3/20}$) and an additional factor of $\sqrt{2}$ which arises from summing over two independent channels, which we want to preserve in our SNR estimate.

The signal to noise described in this work is simply an estimate which uses a time averaged approximation, and is not a substitute for a full time-dependent \emph{LISA} simulation. Our estimates do account for sky location and inclination, as these parameters are well constrained for the sources in this work, and do have significant impacts on the estimated signal to noise. For the purposes of our SNR estimates, we marginalized over the polarization angle. We repeated our calculations using the waveforms given in \citet{Robson2019}, and arrived at the same estimated signal to noise ratios using these combined with the orbit averaged signal (we compute the orbital averaged response in Equation \ref{eq:LISA_AMP}, using the waveforms given by \citet{Robson2019} Equation 15, with the amplitude given by \citet{Robson2019} Equation 20, and use this quantity as the expectation value of the numerator in \citet{Robson2019} Equation 36, and arrive at the same result as the formalism outlined above).

Combined with a distance and inclination estimate, knowing the chirp mass allows for a direct estimate of the gravitational wave strain of the source as measured from Earth. We report the estimated \emph{LISA} SNR for the sources in the sample in Table \ref{tab:LISA}.

It is worth noting that these SNRs are a factor of $\sqrt{2}$ different than those presented in \citet{Korol2017,Burdge2019a,Burdge2019b} due to an error which introduced a factor of two into the strain amplitude, in combination with a $\sqrt{\frac{1}{2}}$ from the prefactor of $\frac{1}{2}$ we omitted in Equation \ref{eq:GWAmp} in this work (which was not omitted in previous work). We also used the more up to date \emph{LISA} sensitivity given in \citet{Robson2019}, which changed results at the $\sim 10 \%$ level compared to the sensitivity of \citet{Amaro-Seoane2017}.

\end{widetext}


\end{document}